\documentclass[12pt]{article}
\usepackage{epsfig}

\newcommand{\lsim}{\raisebox{-4pt}{$\,\stackrel{\textstyle
                                                         <}{\sim}\,$}}

\newcommand{\da}{{distribution amplitude}}
\def\qb{\overline{Q}}
\def\qqb{\overline{Q}\hspace{1pt}^2}
\def\als{\alpha_s}
\def\mev{\,{\rm MeV}}
\def\gev{\,{\rm GeV}}
\newcommand{\qbq}{{q \overline{q}}}
\newcommand{\ubu}{{u \overline{u}}}
\newcommand{\dbd}{{d \overline{d}}}
\newcommand{\sbs}{{s \overline{s}}}
\newcommand{\glg}{{gg}}

\begin{document}
\thispagestyle{empty}
\begin{flushright}
WU B 01-03 \\
DESY-01-118 \\
hep-ph/0108220\\
August 2001\\[5em]
\end{flushright}

\begin{center}
\end{center}
\begin{center}

{\Large\bf The annihilation of virtual photons \\[0.5em]
into pseudoscalar mesons} \\
\vskip 3\baselineskip

M.\ Diehl\,\footnote{Email: markus.diehl@desy.de  }
\\[0.5em]
{\small {\it Deutsches Elektronen-Synchrotron DESY, 22603 Hamburg,
Germany}}\\
\vskip 2\baselineskip

P.\ Kroll\,\footnote{Email: kroll@theorie.physik.uni-wuppertal.de}
and C.\ Vogt\,\footnote{Email: cvogt@theorie.physik.uni-wuppertal.de}
\\[0.5em]
{\small {\it Fachbereich Physik, Universit\"at Wuppertal, 42097
Wuppertal, Germany}}\\

\vskip \baselineskip

\end{center}

\vskip 3\baselineskip
\begin{abstract}
We investigate the possibility to constrain the pion distribution
amplitude from the $\gamma^* \gamma^* \to\pi$ transition. For a
surprisingly large range in the two photon virtualities we find that
the transition form factor is essentially independent of the distribution
amplitude. This in turn entails a parameter-free prediction of
QCD. The $\gamma^* \gamma^* \to\eta,\, \eta'$ form factors are also
briefly discussed. We estimate that experimental studies might be
feasible at the existing $e^+e^-$ experiments BaBar, Belle, and CLEO.
\end{abstract}

\newpage

%%%%%%%%%%%%%%%%%%%%%%%%%%%%%%%%%%%%%%%%%%%%%%%%%%%%%%%%%%%%%%%%%%%%
\section{Introduction}
%%%%%%%%%%%%%%%%%%%%%%%%%%%%%%%%%%%%%%%%%%%%%%%%%%%%%%%%%%%%%%%%%%%%

One of the simplest exclusive observables is the form factor
$F_{P\gamma^{(*)}}$ for transitions from a real or virtual photon to a
pseudoscalar meson $P$. Its behavior at large momentum transfer is
determined by the expansion of a product of two electromagnetic
currents about light-like distances \cite{lep79}. The form factor then
factorizes into a hard scattering amplitude, which is known including
the first-order perturbative QCD corrections \cite{agu81,bra83}, and a
soft matrix element, parameterized by a process independent meson
\da{} $\Phi(\xi)$.

For space-like momentum transfer the form factors $F_{P\gamma^{(*)}}$
can be accessed in $e^+ e^- \to e^+ e^- P$. The measurement by CLEO
\cite{cleo98} for quasi-real photons and $P=\pi$, $\eta$, $\eta'$ has
renewed the interest in these quantities, and many papers have been
devoted to their theoretical analysis, e.g.\
\cite{jak96}--\cite{mel01}, to name a few. The CLEO data are
consistent with \da{}s of the pion, the $\eta$, and the $\eta'$ which
are rather close to the asymptotic form,
\begin{equation}
\Phi_{\rm AS}(\xi)= \frac{3}{2} (1-\xi^2) ,
\end{equation}
where $\xi=2x-1$, and $x$ is the usual momentum fraction carried by
the quark inside the meson.

The purpose of the present article is to investigate the information
contained in the form factors for $\gamma^* \gamma^* \to P$
transitions, beyond what we have already learned from the real-photon
case. In Sect.~\ref{sec:leading-twist} we cast the leading twist,
next-to-leading order result for the $\gamma^* \gamma^* \to\pi$ form
factor in a form useful for the purpose of our study. In
Sect.~\ref{sec:real} we attempt a critical appraisal of what we can
and what we cannot deduce from the existing data on the transition
$\gamma^* \gamma \to\pi$.  The following two sections explore the
$\gamma^*$--$\,\pi$ transition form factor in two different
kinematical regimes. In Sect.~\ref{sec:eta} we briefly point out the
specifics of the transitions to $\eta$ and $\eta'$ mesons. Estimates
of cross sections at the running experiments BaBar, Belle, and CLEO
are given in Sect.~\ref{sec:cross}, and we conclude in
Sect.~\ref{sec:sum}. Some technical details concerning the $\als$
corrections to $F_{\pi\gamma^*}$ are given in an appendix.

%%%%%%%%%%%%%%%%%%%%%%%%%%%%%%%%%%%%%%%%%%%%%%%%%%%%%%%%%%%%%%%%%%%%%%%
\section{The $\gamma^*$--$\,\pi$ transition form factor to leading twist}
\label{sec:leading-twist}
%%%%%%%%%%%%%%%%%%%%%%%%%%%%%%%%%%%%%%%%%%%%%%%%%%%%%%%%%%%%%%%%%%%%%%%

Let us begin with the discussion of the $\gamma^* \gamma^* \to\pi$
form factor.  The $\gamma^*\gamma^*\pi$ vertex is parameterized by
\begin{equation}
\Gamma_{\mu\nu} = - i e^2\, F_{\pi\gamma^*} (\qb,\omega)\,
\varepsilon_{\mu\nu\alpha\beta}\, q^\alpha q'^\beta\, ,
\end{equation}
where we use the convention $\epsilon_{0123}= 1$. Here $q$ and $q'$
respectively denote the photon momenta corresponding to the Lorentz
indices $\mu$ and $\nu$. We introduce the spacelike photon
virtualities $Q^2=-q^2$, $Q'^2=-q'^2$, as well as
\begin{equation}
\qqb = \frac{1}{2}\, (Q^2 + Q'^2) , \qquad
\omega= \frac{Q^2 - Q'^2}{Q^2 + Q'^2} .
\end{equation}
The values of $\omega$ range from $-1$ to $1$, but due to Bose
symmetry the transition form factor is symmetric in this variable:
$F_{\pi\gamma^*}(\qb,\omega) = F_{\pi\gamma^*}(\qb,-\omega)$.

\begin{figure}
\begin{center}
\leavevmode
\psfig{figure=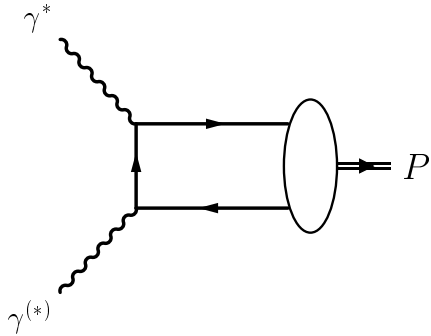,width=5.7cm}
\end{center}
\caption{\label{fig:pigafg}Lowest order Feynman graph for the
$\gamma^* \gamma^{(*)} \to\pi$ transition. A second graph is obtained
by interchanging the photon vertices.}
\end{figure}

To leading-twist accuracy, i.e., in the collinear approximation and
using only the valence Fock state of the pion, the transition form
factor $F_{\pi\gamma^*}$ reads~\cite{agu81,bra83}
\begin{equation}
F_{\pi\gamma^*}(\qb,\omega) = 
\frac{f_\pi}{3\sqrt{2}\, \qqb}\,
\int_{-1}^{\;1} {d} \xi\, \frac{\Phi_\pi(\xi,\mu_F)}{1-\xi^2\omega^2}\, 
\left[1 + \frac{\als(\mu_R)}{\pi}\,{\cal K}(\omega,\xi,\qb/\mu_F) 
\right] \,.
\label{fpgvirtual}
\end{equation}
The Feynman graphs contributing to leading order (LO) are shown in
Fig.~\ref{fig:pigafg}, and the next-to leading order (NLO) kernel
${\cal K}(\omega,\xi,\qb/\mu_F)$ in the $\overline{\rm MS}$ scheme is
given in the appendix.  $\mu_F$ and $\mu_R$ respectively denote the
factorization and renormalization scales, both to be taken of order
$\overline{Q}$, and $f_\pi\approx 131 \mev$ is the well-known pion
decay constant. $\Phi_\pi$ is 
the pion \da, which we expand upon Gegenbauer polynomials
$C_n^{3/2}(\xi)$, the eigenfunctions of the leading-order evolution
kernel for mesons~\cite{lep79},
\begin{equation}
\Phi_{\pi}(\xi,\mu_F)=\Phi_{\rm AS}(\xi)
\left[1+ \sum^\infty_{n=2,4,...}B_n (\mu_F)
\,C_n^{3/2}(\xi)\right] \,.
\label{evoleq}
\end{equation}
The Gegenbauer coefficients are scale dependent, to LO they evolve
according to
\begin{equation}
B_n (\mu_F)= B_n (\mu_0)\,
  \left(\frac{\als(\mu_F)}{\als(\mu_0)}\right)^{\gamma_n /\beta_0}\,,
\label{evolve}
\end{equation}
where $\mu_0$ is the starting scale of evolution and $\beta_0 = 11 -
2n_f /3$.  The anomalous dimensions $\gamma_n$ are positive numbers
increasing with $n$, so that for $\ln{\mu_F}\to\infty$ any \da{}
evolves into $\Phi_{\rm AS}$ with higher order terms in (\ref{evoleq})
gradually becoming suppressed. Hence, the limiting behavior of the
transition form factor for real photons is
\begin{equation}
F_{\pi\gamma}(\qb,\omega=\pm 1) \longrightarrow  \frac{f_{\pi}}
{\sqrt{2}\, \qqb} \,,
\label{asy}
\end{equation}
which is a parameter-free QCD prediction once $f_\pi$ is known. Note,
however, that this limit is only approached logarithmically, and that
the anomalous dimensions are not very large at small $n$. For $n_f=4$
flavors one has $\gamma_2 /\beta_0 = 2/3$, $\gamma_4/\beta_0 \approx
0.97$, $\gamma_6 /\beta_0 \approx 1.17$. As $n$ becomes large the
$\gamma_n$ grow logarithmically, $\gamma_n\sim \frac{16}{3} \ln(n+1)$
being a good approximation already for $n=2$.

To NLO accuracy, the $C_n^{3/2}(\xi)$ are no longer eigenfunctions of
the evolution, so that their coefficients do no evolve
independently. Namely, $B_n(\mu_F)$ at $\mu_F>\mu_0$ depends on all
coefficients $B_2(\mu_0)$, \ldots, $B_n(\mu_0)$.  NLO evolution resums
logarithms $\als^2 \log(\mu_F/\mu_0)$, compared with $\als
\log(\mu_F/\mu_0)$ in LO evolution, and its effects will be more
important when one evolves over a large interval in $\mu_F$ or when
$\als$ at the starting scale $\mu_0$ is large.

Using the Gegenbauer expansion (\ref{evoleq}), the integral in
(\ref{fpgvirtual}) can be worked out analytically order by order in
$n$, provided that $\mu_R$ is chosen to be independent of $\xi$. This
results in
\begin{equation}
F_{\pi\gamma^*}(\qb,\omega) = \frac{f_{\pi}}{\sqrt{2}\: \qqb} 
 \left[ c_0(\omega,\mu_R) 
 + \sum_{n=2,4,\ldots} c_n(\omega,\mu_R,\qb/\mu_F)\, B_n(\mu_F)
\right] \,,
\label{f-two}
\end{equation}
where the lowest-order coefficients $c_n$ read
\begin{eqnarray}
c_0 &=&
\frac{1}{\omega^2} 
  \left[ 1 - (1-\omega^2)\, \frac{\mathrm{artanh}\,\omega}{\omega}
\right] 
- \frac{ \als(\mu_R)}{\pi}\, \frac{1}{9\,\omega^2} 
  \left[ 15 - (1-\omega^2)\, (15 + 4\,\mathrm{artanh}^2\omega)
              \frac{\mathrm{artanh}\,\omega}{\omega} \right] ,
\nonumber\\ 
c_2 &=&
\frac{1}{2\,\omega^4} 
  \left[ 15 - 13\,\omega^2 - (5 - 6\,\omega^2 + \omega^4)\,
              \frac{3\,\mathrm{artanh}\,\omega}{\omega} \right]
+ \frac{ \als(\mu_R)}{\pi} \,{\cal K}_2 (\omega,\qb/\mu_F) ,
\nonumber\\ 
c_4 &=&
\frac{1}{8\,\omega^6} 
  \left[ 315 - 420\,\omega^2 + 113\,\omega^4 
         - (21 - 35\,\omega^2 + 15\,\omega^4 - \omega^6)\,
           \frac{15\,\mathrm{artanh}\,\omega}{\omega} \right]
\nonumber\\ 
&& {}+ \frac{ \als(\mu_R)}{\pi} \,{\cal K}_4 (\omega,\qb/\mu_F) .
\label{two}
\end{eqnarray}
The analytical expressions of the $\als$ corrections ${\cal K}_2$ and
${\cal K}_4$ are rather lengthy and we refrain from showing them
explicitly. Their dependence on $\ln(\qb/\mu_F)$ is partially
compensated in $F_{\pi\gamma^*}(\qb,\omega)$ by the $\mu_F$ dependence
of the $B_n(\mu_F)$. Notice that no such compensation takes place for
the $\mu_R$ dependence to the order in $\als$ we are working
in. Unless stated otherwise we will in the following take
$\mu_F=\mu_R=\qb$, which is the virtuality of the quark propagators in
Fig.~\ref{fig:pigafg} at $\xi=0$. The coefficients $c_n$ then depend
weakly on $\qb$ via $\als(\qb)$.  Other scale choices lead, as usual,
to results differing by terms of ${\cal O}(\als^2)$, which is beyond
the accuracy of our analysis.  We finally remark that for fixed
$\omega$ the $F_{\pi\gamma^*}$ form factor only falls off like
$\qb{}^{-2}$ at large $\qb$, in contrast to the $Q^{-2}Q'{}^{-2}
\propto \qb{}^{-4}$ behavior of the vector meson dominance
model~\cite{kes93}.

\begin{figure}
\begin{center}
\epsfig{file=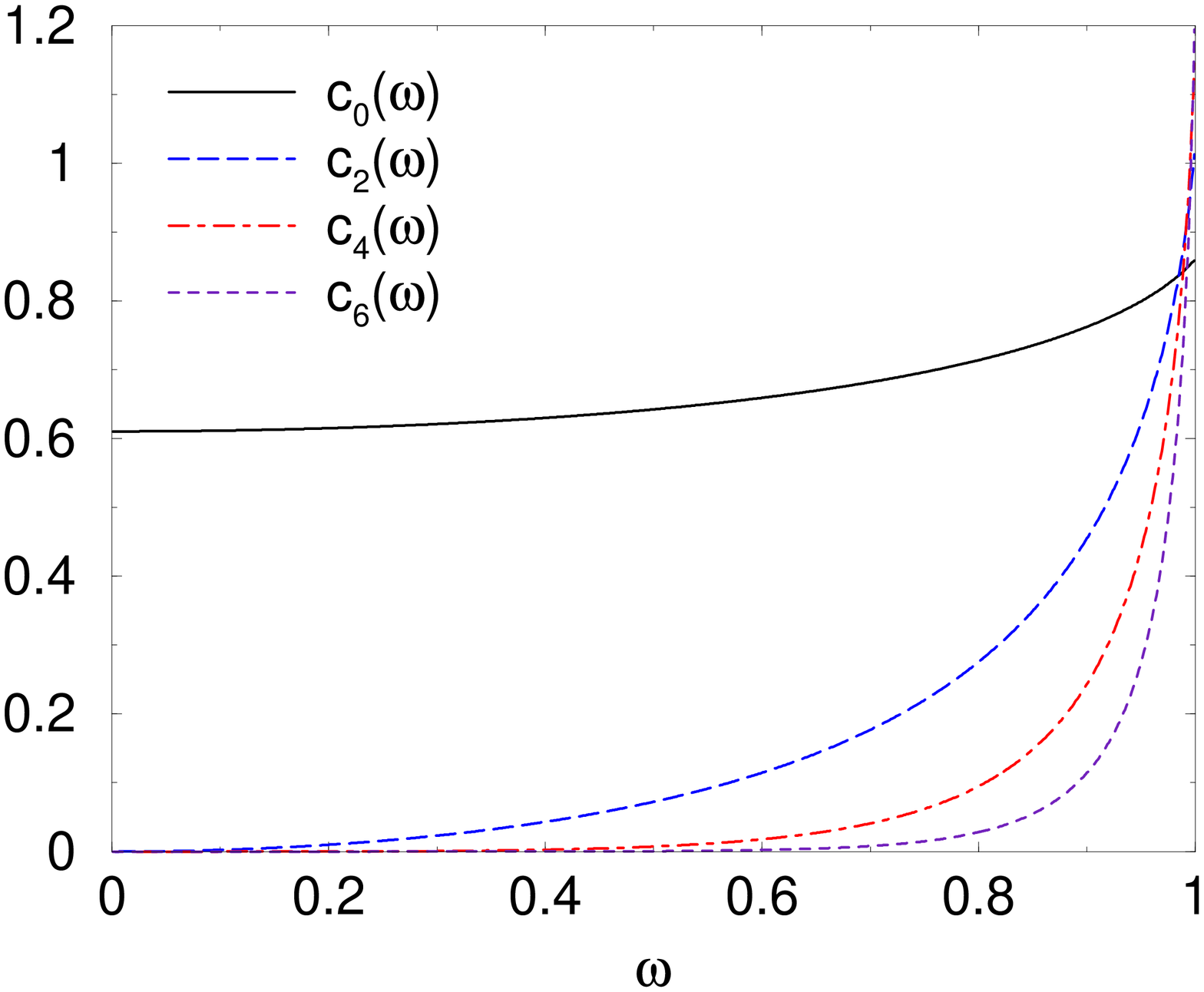,bb=90 70 560 665,width=5.5cm,angle=-90}
\hspace{1cm}
\epsfig{file=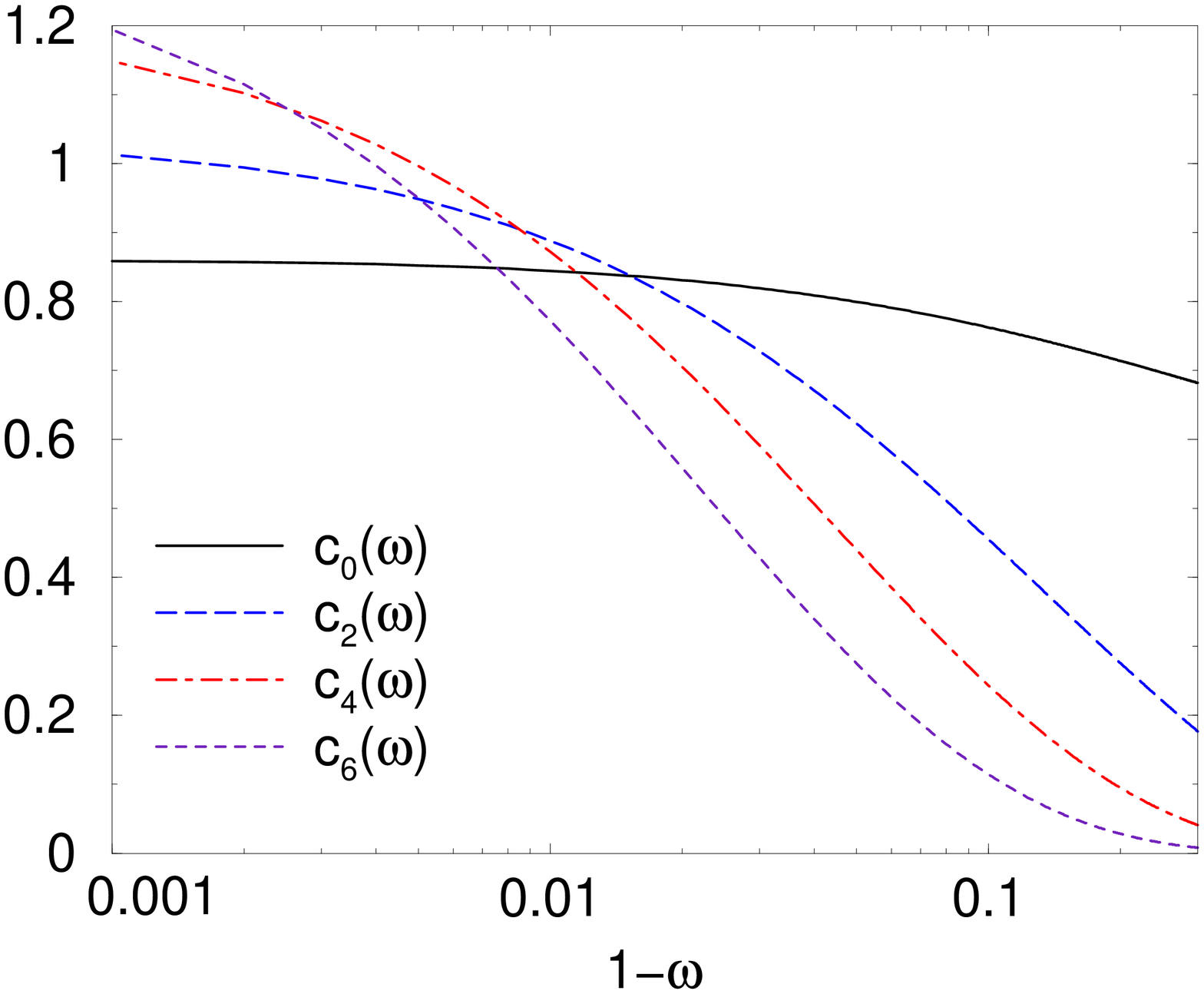,bb=90 30 560 665,width=5.5cm,angle=-90}
\end{center}
\caption{\label{fig:coeff} The coefficients $c_n(\omega)$ in the
expansion (\protect\ref{f-two}) of the $\gamma^*$--$\,\pi$ transition
form factor. NLO corrections are included with $\mu_F=\mu_R=\qb$,
which is taken as 2 GeV.}
\end{figure}

In order to visualize the sensitivity of the form factor to the
Gegenbauer coefficients we plot the lowest coefficients $c_n(\omega)$
in Fig.~\ref{fig:coeff}. Here and in the following we use the two-loop
expression of $\als$ for $n_f=4$ flavors and
$\Lambda^{(4)}_{\overline{\rm MS}}= 305 \mev$ \cite{cas98}. We see a
surprising behavior of the coefficients in $\omega$. In the
real-photon limit $\omega\to 1$ the form factor is sensitive to all
Gegenbauer moments with approximately equal weight, $c_n(\omega=1)
\approx 1$. As soon however as one departs from this limit, the
coefficients $c_n(\omega)$ decrease and become ordered as $c_0 > c_2 >
c_4 > \ldots$\,. Except for $c_0$ this decrease is rather fast. If
$\omega < 0.8$, for instance, the second coefficient $c_2$ is less
than 40\% of $c_0$.  For a wide range of $\omega$, the form factor is
essentially independent of the $B_n$, unless they are unexpectedly
large. The discussion of what can be learned from the measurement of
$F_{\pi\gamma^*}$ naturally falls into two parts, concerning the
kinematic regions $Q'^2 \ll Q^2$ and $Q'^2 \sim Q^2$, respectively.

%%%%%%%%%%%%%%%%%%%%%%%%%%%%%%%%%%%%%%%%%%%%%%%%%%%%%%%%%%%%%%%%%%%%%%%
\section{The real-photon limit}
\label{sec:real}
%%%%%%%%%%%%%%%%%%%%%%%%%%%%%%%%%%%%%%%%%%%%%%%%%%%%%%%%%%%%%%%%%%%%%%%

Before embarking on the study of $F_{\pi\gamma^*}$ let us discuss some
issues in the analysis of the real-photon limit. At $\omega=1$ one has
\begin{equation}
c_n(\omega=1) = 1+ 
\frac{\als(\mu_R)}{\pi}\, {\cal K}_n(\omega=1,\qb/\mu_F)\,,
\end{equation}
where the first few coefficients explicitly read
\begin{equation}
{\cal K}_0 = - \frac{5}{3} , 
\hspace{2em}
{\cal K}_2 = \frac{5}{3} \left( \frac{59}{72} 
		- \frac{5}{6}\, \ln\frac{2 \qqb}{\mu_F^2} \right) ,
\hspace{2em}
{\cal K}_4 = \frac{5}{3} \left( \frac{10487}{4500} 
		- \frac{91}{75}\, \ln\frac{2 \qqb}{\mu_F^2} \right) .
\label{nlo-coeff}
\end{equation}
The $\gamma$--$\,\pi$ transition form factor thus approximately probes
the sum $1 + \sum_n B_n$ of Gegenbauer coefficients. Due to evolution
and the running of $\als$ the relative weights of the $B_n(\mu_0)$ in
$F_{\pi\gamma}$ vary with $Q$, but only logarithmically. Extracting
information on the Gegenbauer coefficients beyond their sum hence
requires analyzing the form factor in a sufficiently large range of
$Q$, say $Q_{min} \leq Q \leq Q_{max}$. While $Q_{max}$ is usually set
by the available data, the choice of $Q_{min}$ is an important source
of theoretical uncertainty, since at lower $Q$ power corrections to
the leading twist result can become increasingly important. There are
various approaches to describe such power corrections, all of which
involve further assumptions or parameters.

A different source of uncertainty is that in practice analyses of
$F_{\pi\gamma}$ data are performed with a truncation of the Gegenbauer
series, setting all $B_n=0$ for $n\ge n_0$ 
with some $n_0$. This may be seen as an analog to the determination of
parton distributions, where one typically chooses a functional form of
the parton distributions at the starting scale of evolution and then
fits its parameters to data on inclusive processes. Notice that
setting $B_n$ to zero for \emph{all} values of $\mu_F$ is strictly
speaking not consistent with NLO evolution, which generates nonzero
$B_n(\mu_F)$ even if $B_n(\mu_0)=0$. M\"uller \cite{mue95} has shown
that this can lead to important effects on the shape of the \da{},
especially in the endpoint regions and when $\als$ at the starting
scale is large. In the analysis \cite{kro96} of the CLEO data, the
impact of NLO evolution on the quantity $F_{\pi\gamma}$ was however
found to be small compared with the NLO corrections to the hard
scattering kernel.  In the numerical studies of the present work we
will be concerned with values of $\mu_F=\qb$ between 1 and 2 GeV. We
choose $\mu_0=1 \gev$ as the starting scale of evolution and take the
LO formula (\ref{evolve}), using however the two-loop expression of
$\als$.  Since the purpose of this paper is not a precise
determination of $\Phi_\pi$, this should be sufficiently accurate. At
the same time it keeps the analysis procedure simple as it allows us
to work with a finite number of Gegenbauer coefficients at all scales.

The simplest analyses of the $F_{\pi\gamma}$ data assume $n_0=4$. This
leads to stable results on $B_2$, although its value is subject to the
uncertainties just discussed. Thus, for instance, Ref.~\cite{kro96}
obtained the value $B_2(\mu_0=1\gev)=-0.15\pm 0.04$ to NLO accuracy in
the $\overline{\rm MS}$ scheme, using $\mu_F=\mu_R =Q$ and
$Q^2_{min}=3\gev^2$. Changing to $\mu_F=\mu_R =\qb$ as we prefer here,
one finds $B_2(\mu_0=1\gev)=-0.06\pm 0.03$ for both
$Q^2_{min}=2\gev^2$ and $3\gev^2$ within errors. This shift in the
value of $B_2$ may be taken as an indication of the uncertainties due
to uncalculated higher orders in the perturbative expansion.  Meli\'c
et al.~\cite{mel01} have calculated the part of the ${\cal
O}(\alpha_s^2)$ corrections that allows one to determine the BLM scale
for $F_{\pi\gamma}$. Taking the \da{} $\Phi_{\rm AS}$ they find $\mu_R
\approx Q/3$ in the $\overline{\rm MS}$ scheme. The corresponding NLO
prediction for $F_{\pi\gamma}$ is slightly below the CLEO data, in
contrast to the one for $\Phi_{\rm AS}$ and $\mu_R=Q$, which is
slightly above.  Brodsky et al.~\cite{bro98} find consistency with the
CLEO data when taking $\Phi_{\rm AS}$ and a yet lower renormalization
scale together with a prescription for the saturation of $\alpha_s$,
thus including effects beyond a leading twist perturbative analysis.

If one allows for $B_2$ and $B_4$ in the analysis there is no unique
result for the individual coefficients. Rather there is a linear
correlation between $B_2$ and $B_4$, which using (\ref{evolve}),
(\ref{f-two}), (\ref{nlo-coeff}) one can approximate as
\begin{eqnarray}
f(\qb) &=& B_2(\mu_0) +
  \left(1 + \frac{\als(\mu_R)}{\pi}\, ({\cal K}_4 - {\cal K}_2)\right) 
  \left(\frac{\als(\mu_F)}{\als(\mu_0)}\right)^{
                                  (\gamma_4 - \gamma_2)/\beta_0}
  B_4(\mu_0) + {\cal O}(\als^2)
\nonumber \\
&\approx& B_2(\mu_0) + \Big( 1 + 0.66 \als(\qb) \Big) 
  \left(\frac{\als(\qb)}
             {\als(\mu_0)}\right)^{0.3} B_4(\mu_0) + {\cal O}(\als^2) ,
\label{corr}
\end{eqnarray}
where we have taken $\mu_F=\mu_R=\qb$ and $n_f=4$ when going from the
first to the second line.  The function $f(\qb)$ includes the data on
$F_{\pi\gamma}$ and the term with $c_0$ in (\ref{f-two}). It may also
absorb possible power corrections as far as they are independent of
$\Phi_\pi$. Only the mild logarithmic $Q$ dependence due to evolution
and the running of $\als$ restricts the allowed values of $B_2$ and
$B_4$ to a finite region in parameter space. The factor multiplying
$B_4(\mu_0)$ in (\ref{corr}) varies between 1.31 and 1.05 for $\qqb$
between 1 and $4 \gev^2$, the latter corresponding to the center of
the highest $Q^2$ bin in the CLEO measurement.

\begin{figure}
\parbox{\textwidth}
{\begin{center}
\psfig{file=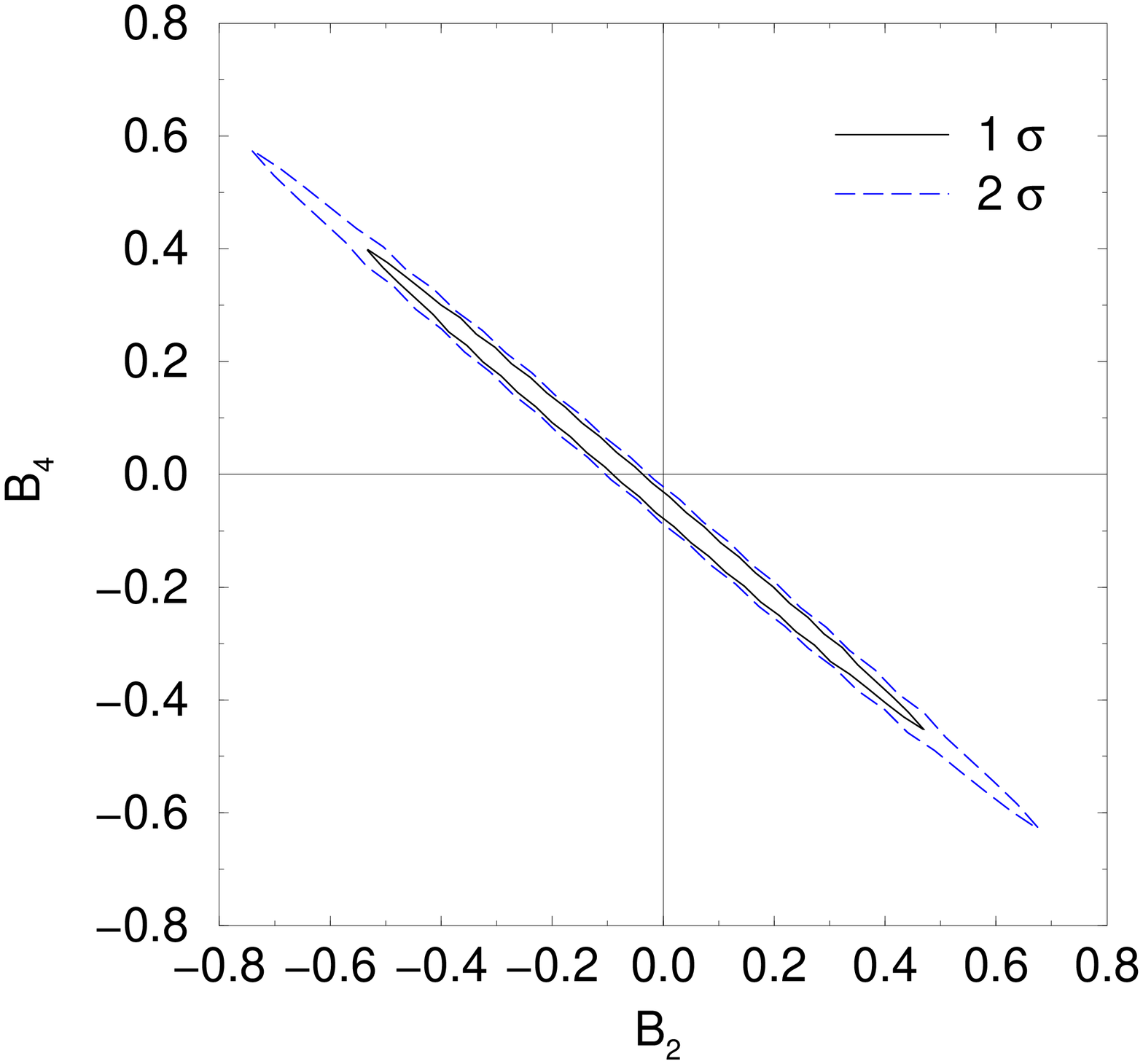,bb=95 55 560 640,width=6cm,angle=-90}
\hspace{1cm}
\psfig{file=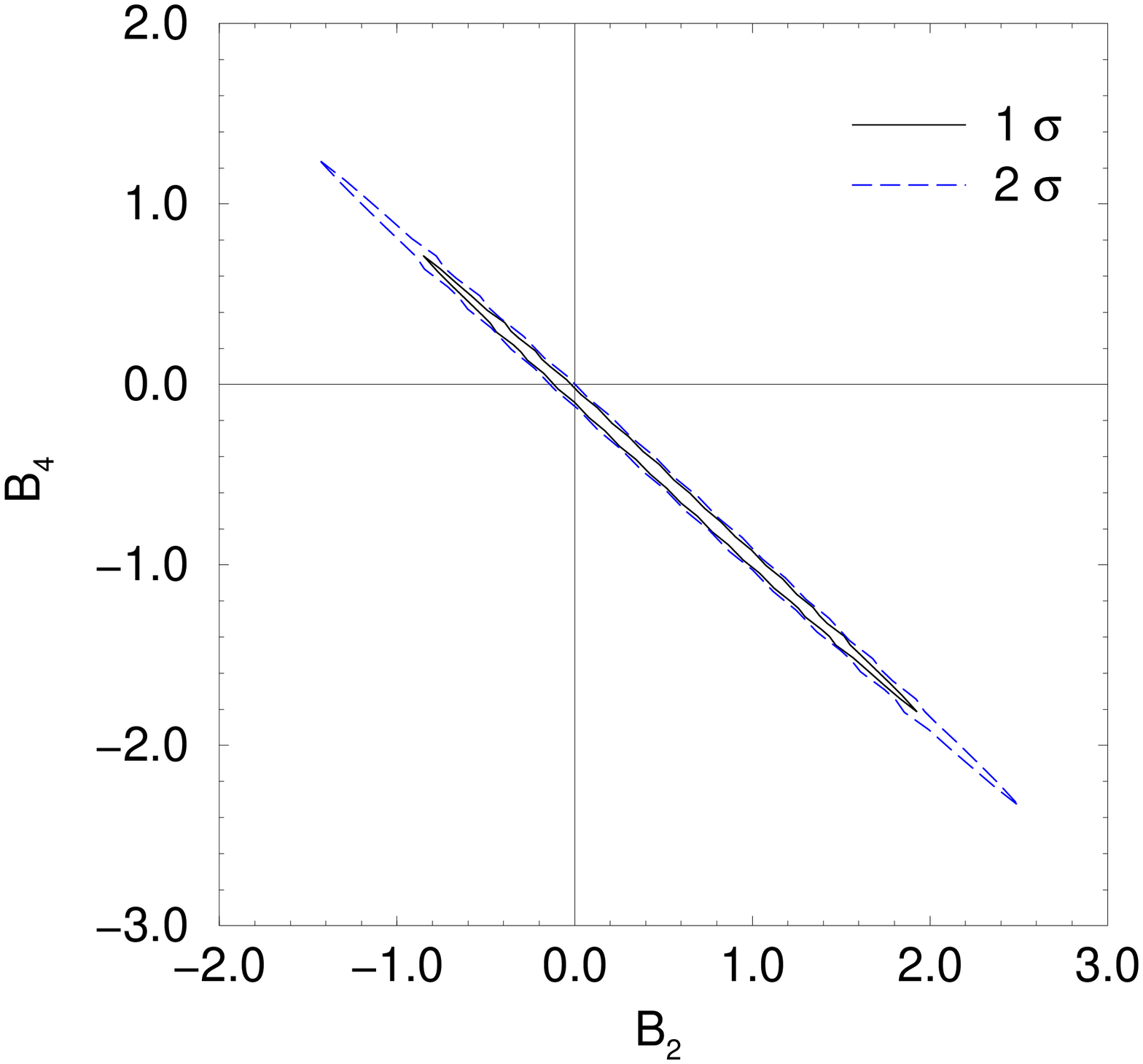,bb=95 55 560 640,width=6cm,angle=-90}
\end{center}
}
\caption{{}$1\,\sigma$ and $2\,\sigma$ $\chi^2$-contour plots for a
two parameter fit to the CLEO data on $F_{\pi\gamma}$ choosing
$Q^2_{min}=2\gev^2$ (left) and $3\gev^2$ (right). Values of the
Gegenbauer coefficients $B_2$ and $B_4$ refer to a factorization scale
of $\mu_0=1 \gev$. The data is fitted to the leading twist NLO
expression (\protect\ref{fpgvirtual}) with
$\mu_F=\mu_R=Q/\sqrt{2}$. The $\chi^2$ is calculated from the errors
on the data and does not include a theory error.}
\label{fig:contour}
\end{figure}

The experimental errors of the form factor data allow deviations from
the linear correlation~(\ref{corr}). This correlation is nicely
illustrated by the $\chi^2$-contour plots in Fig.~\ref{fig:contour},
generated by MINUIT for our leading-twist NLO analysis. Comparison of 
the plots reveals that for $Q^2_{min}=3 \gev^2$ the allowed parameter
regions are enlarged since both the number of data points included in
the fit and the lever arm in $Q$ are smaller. On the other hand, power
corrections should be less important in this case. We also observe
from the figure that the $1\sigma$ range for $B_2$ and $B_4$
obtained from the fit with $Q^2_{min}=2 \gev^2$ is embedded in that
for $Q^2_{min}=3 \gev^2$, indicating that the fits are consistent with
each other. Within the experimental errors, logarithmic effects
suffice to describe the residual $Q^2$ dependence of the CLEO data for
$Q^2 F_{\pi\gamma}(Q^2)$ above $2 \gev^2$.  We emphasize that this
finding does not \emph{prove} that power corrections are indeed small
in that region, it rather illustrates the difficulty to distinguish a
power from a logarithmic behavior in $Q^2$ with data in the range
between 2 and $8 \gev^2$.

{}From the above exercise we conclude that only extreme values of
$|B_2|$ and $|B_4|$, say above 1 or 2, are ruled out if higher order
Gegenbauer coefficients are neglected.  A compact way of presenting
the result of the fit with two free coefficients is to use the linear
combinations $B_2 + B_4$ and $B_2 - B_4$, which have approximately
uncorrelated errors. Taking $Q^2_{min} =2 \gev^2$ we obtain $B_2+B_4 =
- 0.06 \pm 0.08$ and $B_2-B_4 = 0.0 \pm 0.9$ at $\mu_0=1 \gev$.  This
illustrates that, within a leading twist NLO analysis, the CLEO data
on the $\gamma^* \gamma \to\pi$ form factor is insufficient for an
unambiguous determination of the pion \da, rather it approximately
fixes the sum $\sum_n B_n$ to be close to zero. We cannot decide from
the existing data on $F_{\pi\gamma}$ whether the small value of the
sum $\sum_n B_n$ results from the cancellation of rather large
individual terms or from the smallness of the $B_n$
themselves. Additional information from processes where $\Phi_\pi$
enters in a different way, would be highly welcome to settle this
issue. One candidate are the decays of charmonium states into pion
pairs, where however the quality of the data as well as unsuppressed
color octet contributions \cite{schu96} do not permit a conclusive
analysis at present. Many theoretical studies have been devoted to the
elastic pion form factor, cf.\
e.g.~\cite{bro98,bra00,ste00,li92,jak93}, but data at large $Q^2$ is
scarce. A different way to probe the pion wave function is provided by
diffractive dissociation of a pion into jets \cite{ber81}. It is
however not clear at present to which precision information on
$\Phi_\pi$ can be obtained in that process \cite{nik01}.

We will investigate below what data on the $\gamma^*$--$\,\pi$
transition form factor could contribute, but before this we should
discuss the question of theoretical uncertainties in
$F_{\pi\gamma}$. The $\chi^2$-contours in Fig.~\ref{fig:contour} only
reflect the errors on the form factor data. We have already mentioned
the uncertainty from the choice of factorization and renormalization
scales. Much harder to estimate is the role of power
corrections. Contributions from various dynamical sources have been
studied in the literature, such as higher twist distribution
amplitudes \cite{kho99,yeh01}, the nonperturbative behavior of
$\als(\mu)$ in the infrared region \cite{bro98,ste00}, or transverse
momentum effects in the hard scattering subprocess
\cite{jak96,kro96,mus97}.

At this point we wish to comment on the analysis of the CLEO data
presented in \cite{schm00}. Power corrections due to the hadronic
component of the photon have been modeled there within QCD sum
rules. The relative weights of $B_2(\mu_0)$ and $B_4(\mu_0)$ in
$F_{\pi\gamma}$ then display a much stronger $Q^2$ dependence than in
the leading twist case (\ref{corr}), which leads to a much smaller
allowed parameter region than in our Fig.~\ref{fig:contour}. A large
part of the deviation between the $F_{\pi\gamma}$ data and the result
obtained with $\Phi_{\rm AS}$ in that analysis is due to the inclusion
of twist-four \da{}s. Their shape is taken to be the asymptotic one,
and for their normalization results from QCD sum rule calculations as
given in \cite{kho99} are used. No error is however assigned to this
input in the analysis. Note that an uncertainty of about $\pm 30\%$
has been estimated for the relevant normalization constant $\delta^2$
in~\cite{kho01}.  Given this and the fact that only certain power
corrections are taken into account in the QCD sum rule technique, we
feel that the errors on the extracted Gegenbauer coefficients given in
\cite{schm00} are subject to a significant model dependence.

Several of the analyses cited above find moderate but non-negligible
power corrections to $F_{\pi\gamma}$ in the $Q^2$ range where most of
the CLEO data is concentrated. We find that our limited ability to
reliably describe, let alone to calculate power corrections precludes
us from drawing firm quantitative conclusions and from discriminating
between many of the available theory predictions for $\Phi_\pi$. The
best remedy to this situation we can see is data on $F_{\pi\gamma}$ at
the highest possible $Q^2$. Given the luminosity of the presently
running experiments BaBar, Belle and CLEO, a substantial improvement
should be possible over the statistics of the available data
\cite{cleo98}, which is based on an integrated luminosity of about 3
fb$^{-1}$.

Notice that the range of theory predictions on $\Phi_\pi$ is
considerable. For instance, the QCD sum rule analysis of Braun and
Filyanov \cite{bra89} gave $B_2(1~\mbox{GeV})=0.44$ and
$B_4(1~\mbox{GeV})=0.25$ under the assumption that higher order
coefficients are negligible. On the other side, a still preliminary
result from lattice QCD \cite{lat99} gives $B_2= - 0.41\pm 0.06$ at a
low scale. Between these extremes many studies, using e.g.\ light-cone
QCD sum rules with non-local condensates \cite{bak01}, the transverse
lattice \cite{dal01}, or the instanton model of QCD \cite{goe98}
obtain a \da{} either slightly broader or slightly narrower than
$\Phi_{\rm AS}$.

%%%%%%%%%%%%%%%%%%%%%%%%%%%%%%%%%%%%%%%%%%%%%%%%%%%%%%%%%%%%%%%%%%%%%%%
\section{The region $Q'^2 \ll Q^2$}
\label{sec:asymmetric}
%%%%%%%%%%%%%%%%%%%%%%%%%%%%%%%%%%%%%%%%%%%%%%%%%%%%%%%%%%%%%%%%%%%%%%%

With the lessons from the real-photon limit in mind let us now
investigate the region where one of the photons is slightly
off-shell. In order to gain some insight in the importance of power
corrections we estimate transverse momentum effects by employing the
modified perturbative approach \cite{bot89,li92}. It has been applied
to the case of $\gamma^*$--$\,\pi$ transitions in \cite{jak96,kro96},
and to the case at hand in \cite{ong95}. In this approach the
expression (\ref{fpgvirtual}) is replaced by
\begin{equation}
F_{\pi\gamma^*}(\qb,\omega) = \frac{1}{4\sqrt{3}\pi^2} \int {d} \xi\,
     {d}^2{\bf b}\, \hat\Psi_{\pi}^*(\xi,-{\bf b},\mu_F)\,
K_0(\sqrt{1 + \xi \omega}\: \qb\, b)\, 
     \exp\left[-S\left(\xi,b,\qb,\mu_R\right)\right]\,,
\label{fpgeq}
\end{equation}
where $K_0$ is the modified Bessel function of order zero,
representing the Fourier transform of the leading-order hard
scattering amplitude in momentum space. The quark-antiquark separation
${\bf b}$ is canonically conjugated to the usual transverse momentum
${\bf k}_{\perp}$. The Sudakov exponent $S$ describes gluonic
radiative corrections not taken into account in the evolution of the
wave function. For $\ln\qb \to\infty$ it suppresses all contributions
to the integral except for those with small quark-antiquark
separations. As a consequence the limiting behavior (\ref{asy})
emerges for $\gamma$--$\,\pi$ transitions, as we have checked
numerically by calculating expression~(\ref{fpgeq}) up to $Q^2=10^{15}
\gev^2$.  As $b$ sets the interface between non-perturbative soft
gluons contained in the hadronic wave function and perturbative soft
gluon contributions resummed in the Sudakov factor, the factorization
scale $\mu_F$ is taken as $1/b$.  For the renormalization scale we
take the prescription $\mu_R = \max{\{1/b, \sqrt{1+\xi\omega}\:\, \qb,
\sqrt{1-\xi\omega}\:\, \qb\}}$ of \cite{li92}. Following
\cite{lep83,jak93} we take for the light-cone wave function in
$b$-space the simple form
\begin{equation}
\hat{\Psi}_\pi(\xi,{\bf b})=\frac{2\pi f_\pi}{\sqrt{6}}\, 
       \Phi_{\rm AS}(\xi) 
       \exp{\left[-\frac{ \pi^2 f_\pi^2}{2} (1-\xi^2)\right]}
\label{modwf}
\end{equation}
in our estimate. Evaluation of the $\gamma^* \gamma \to\pi$ form
factor in the modified perturbative approach using this wave function
leads to very good agreement with the CLEO data \cite{kro96}.

\begin{figure}
\begin{center}
\leavevmode
\epsfxsize=0.5\textwidth  \epsffile{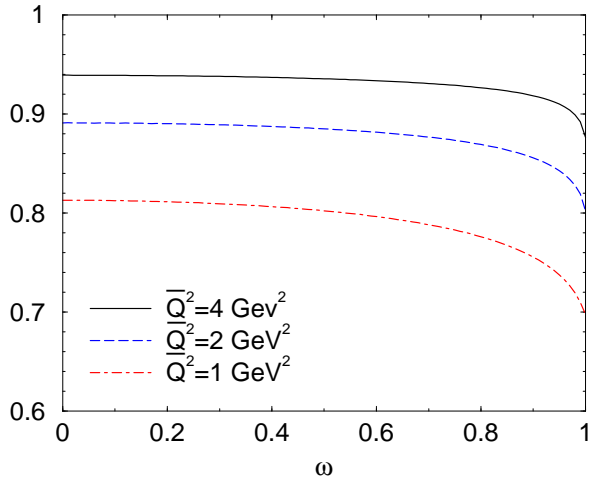}  
\end{center}
\caption{Ratio of $F_{\pi\gamma^*}(\qb,\omega)$ in the modified
perturbative approach and in the LO leading twist approximation at
$\qqb= 4\gev^2$ (solid line), $\qqb=2\gev^2$ (dashed line), and $\qqb=
1\gev^2$ (dash-dotted line). Here we have used the wave function
(\ref{modwf}) in the modified perturbative approach and the asymptotic
pion \da{} in the leading twist calculation.}
\label{fig:ratio} 
\end{figure}

In the kinematical range of interest here we find that the Sudakov
factor only provides corrections of no more than $1.5\%$ to
$F_{\pi\gamma^*}$, and it is thus good enough to retain only the
leading logarithmic terms in $S$ as given in \cite{bot89}. In our
kinematics the difference between the asymptotic result (\ref{asy})
and the expression (\ref{fpgeq}) is thus essentially due to the ${\bf
k_\perp}$-corrections to the hard scattering amplitude and not to the
perturbative corrections contained in the Sudakov factor. In order to
estimate the importance of power corrections we therefore compare
(\ref{fpgeq}) with the leading twist result at LO rather than at NLO
in $\als$.  In Fig.~\ref{fig:ratio} we show the ratio between the form
factor evaluated from (\ref{modwf}) in the modified perturbative
approach and the LO result calculated with $\Phi_{\rm AS}$ in the
leading-twist approximation. As we can see, the corrections are below
10\% for values of $\qqb= 4 \gev^2$, but can go up to 30\% for $\qqb =
1 \gev^2$.  As expected, the importance of power corrections decreases
as both photons become virtual. This is already signaled by the
leading twist result. Indeed, the factor factor $1/(1-\xi^2\omega^2)$
in the convolution (\ref{fpgvirtual}) controls to which extent
$F_{\pi\gamma^*}$ is sensitive to contributions from the end-point
regions $\xi \to \pm 1$, where the quark or antiquark in the pion
becomes slow and soft effects can become important.

\begin{figure}
\begin{center}
\leavevmode
\epsfxsize=0.55\textwidth  \epsffile{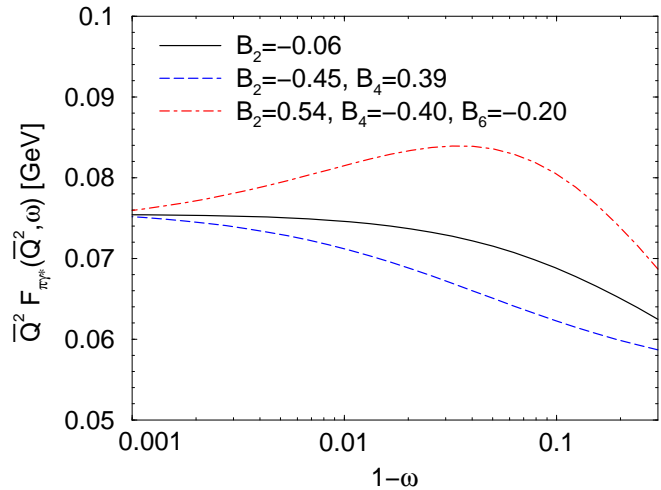}  
\caption{The scaled form factor $\qqb F_{\pi\gamma^*}(\qb,\omega)$
calculated to NLO in the leading-twist approximation at $\qqb=4
\gev^2$, using sample \da{}s. The values of the Gegenbauer
coefficients are quoted at the scale $\mu_0 = 1 \gev$.  The curves are
evaluated from $B_2=-0.06$ (solid line), from $B_2=-0.45$, $B_4=0.39$
(dashed line) and from $B_2=0.54$, $B_4=-0.40$ and $B_6=-0.20$
(dash-dotted line). All higher order Gegenbauer coefficients are taken
to be zero.}
\label{endplot}
\end{center}
\end{figure}

Let us explore how much information on the pion \da{} can be obtained
from the $\omega$ dependence of $F_{\pi\gamma^*}$ under these
circumstances. To this end, we plot in Fig.~\ref{endplot} the form
factor for different choices of \da{}s. These have been chosen to give
the same value of $\sum_n B_n \simeq -0.06$ at the scale
$\mu_0=1\gev$, so that up to small corrections they all give the same
value of $F_{\pi\gamma}$ in the NLO leading-twist analysis. We see
that for Gegenbauer coefficients whose order of magnitude is not
implausibly large compared to the theory estimates we have quoted
above, one can obtain visible differences in the form factor. In part
of the $\omega$ range they can attain 15\% and are thus marginally
above the level where we have estimated that at $\qqb= 4 \gev^2$ power
corrections can make a reliable extraction of the $B_n$
problematic. While one clearly has not enough discriminating power to
pin down individual coefficients $B_n$, one can gain valid information
beyond what can be inferred from real-photon data. Notably, one can
check whether the small value of the sum $\sum_n B_n$, to which
$F_{\pi\gamma}$ is mainly sensitive, results from the cancellation of
rather large individual terms or from the smallness of the $B_n$
themselves. This type of evidence would be rather complementary to the
quantitatively more precise information we have argued to be
accessible from high $Q^2$ data on $F_{\pi\gamma}$.

%%%%%%%%%%%%%%%%%%%%%%%%%%%%%%%%%%%%%%%%%%%%%%%%%%%%%%%%%%%%%%%%%%%%%%%
\section{The region $Q'^2 \sim Q^2$}
\label{sec:symmetric}
%%%%%%%%%%%%%%%%%%%%%%%%%%%%%%%%%%%%%%%%%%%%%%%%%%%%%%%%%%%%%%%%%%%%%%%

As we have seen in Fig.~\ref{fig:coeff}, the contribution from
Gegenbauer coefficients $B_n$ to $F_{\pi\gamma^*}$ becomes small as
one goes away from $\omega=1$, with a faster rate of decrease for
increasing index $n$. To understand this, we observe that the hard
scattering kernel in (\ref{fpgvirtual}) can be Taylor expanded around
$\omega=0$. Using the Gegenbauer expansion (\ref{evoleq}) of
$\Phi_\pi(\xi)$, we find the relevant integral to be
\begin{eqnarray}
\lefteqn{ \int_{-1}^{1} d\xi\, \frac{1-\xi^2}{1-\xi^2\omega^2}
 \left[1+ \sum^\infty_{n=2,4,\ldots} B_n \,C_n^{3/2}(\xi)\right] }
\nonumber \\
&=& \sum^\infty_{m=0,2,\ldots}  \omega^m\, 
                \int_{-1}^{1} d\xi\, (1-\xi^2)\, \xi^m 
+ \sum^\infty_{\scriptstyle n=2,4,\ldots 
           \atop \scriptstyle m=n,n+2,\ldots}  \omega^m B_n \,
    \int_{-1}^{1} d\xi\, (1-\xi^2)\, \xi^m C_n^{3/2}(\xi)
\hspace{1em}
 \label{omega-limit}
\end{eqnarray}
for the LO contribution to $F_{\pi\gamma^*}$. The condition $m\ge n$
in the sum involving the $B_n$ incorporates the orthogonality of
Gegenbauer polynomials, as the $\xi$ integrals for $m<n$ are zero. We
thus have the remarkable result that in the limit $\omega\to 0$ a
Gegenbauer coefficient $B_n$ is suppressed in $F_{\pi\gamma^*}$ by a
power $\omega^n$. This also holds to NLO, as we will show in the
appendix, and explains the behavior of the coefficients $c_n$ in
Fig.~\ref{fig:coeff}. Keeping only terms up to ${\cal O}(\omega^2)$,
we have
\begin{eqnarray}
F_{\pi\gamma^*}(\qb,\omega)&=&\frac{\sqrt{2}f_\pi}{3\: \qqb} \left[
    1 - \frac{\als}{\pi} + 
  \frac15\, \omega^2 \left( 1-\frac53\frac{\als}{\pi} \right)
  \phantom{\frac{\qqb}{\mu_F^2}} \right. 
\nonumber \\
&&  \hspace{2em} \left. {}+ \frac{12}{35}\, \omega^2 B_2(\mu_F) 
    \left( 1 + \frac{5}{12} \frac{\als}{\pi} 
    \left\{ 1-\frac{10}{3} \ln\frac{\qqb}{\mu_F^2} \right\}
    \right) \right]
+ {\cal O}(\omega^4,\als^2)\, .
\label{fpgapprox}
\end{eqnarray}
The limiting behavior for $\omega\to 0$ and to leading order in $\als$
has been derived long ago in \cite{cor66}. The $\als$-correction to
the leading term,
\begin{equation}
F_{\pi\gamma^*}(\qb,\omega)=\frac{\sqrt{2}f_\pi}{3\: \qqb}
         \bigg[ 1-\frac{\als}{\pi} \bigg]+ {\cal O}(\omega^2,\als^2)
\label{qcd-pre}
\end{equation}
has already been given in \cite{agu81}.

\begin{figure}
\begin{center}
\leavevmode
\epsfxsize=0.51\textwidth  \epsffile{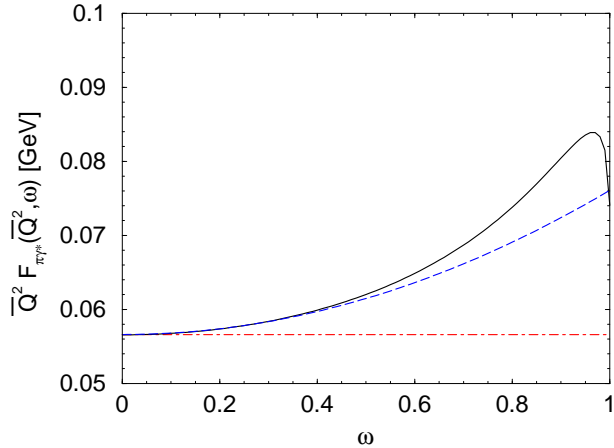}  
\end{center}
\caption{Comparison of a NLO leading twist calculation of the scaled
form factor $\qqb F_{\pi\gamma^*}(\qb,\omega)$ (solid line) with the
approximations~(\ref{fpgapprox}) (dashed line) and~(\ref{qcd-pre})
(dash-dotted line). The form factor is evaluated with
$\mu_F=\mu_R=\qb$ at $\qb=2 \gev$ for a sample \da{} with $B_2=0.54$,
$B_4=-0.40$, $B_6=-0.20$, and $B_n=0$ for $n\geq 8$. Note the
suppressed zero on the $y$ axis.}
\label{fig:pred}
\end{figure}

Given the small numerical coefficients in front of $\omega^2$, the
$\omega$ independent term in Eq.~(\ref{fpgapprox}) dominates over a
rather large range of $\omega$. Even at $\omega\simeq 0.6$ the
$\omega^2$ corrections amount to less than $15\%$ if $|B_2|
<0.5$. Since higher coefficients $B_n$ are suppressed even more
strongly, we conclude that in this range of $\omega$ the
$\gamma^*$--$\,\pi$ transition form factor is essentially flat in
$\omega$ and independent of the pion distribution amplitude
$\Phi_\pi$. To illustrate the quality of the small-$\omega$
approximations we compare in Fig.~\ref{fig:pred} the full result
(\ref{fpgvirtual}) for $F_{\pi\gamma^*}$ with the
expressions~(\ref{fpgapprox}) and~(\ref{qcd-pre}) at $\qb=2\gev$ for a
sample \da{} given by $B_2=0.54$, $B_4=-0.40$, $B_6=-0.20$ at $\mu_0=1
\gev$.  The full calculation is in agreement with the CLEO data for
$\omega\to 1$. We see that, although $B_2$ in our example is quite
large and positive, both approximations are indeed very good for
$\omega \lsim 0.6$.

We thus have a parameter-free prediction of QCD to leading-twist
accuracy, which well deserves experimental verification.  Any observed
deviation from (\ref{qcd-pre}) beyond what can reasonably be ascribed
to ${\cal O}(\als^2)$ terms would be an unambiguous signal for power
corrections. Only if the lowest Gegenbauer coefficients $B_n$ were
extremely large would this conclusion become invalidated, but as we
discussed in the previous section, such a scenario could already be
ruled out using the region $\omega \simeq 1$. We remark that according
to our estimate in Sect.~\ref{sec:asymmetric}, power corrections need
not be negligibly small at moderate values of $\qb$, even for
$\omega=0$.

For small $\omega$, the relation (\ref{qcd-pre}) has a status
comparable to the famous expression of the cross section ratio $R =
\sigma(e^+ e^-\to {\rm hadrons}) / \sigma(e^+e^-\to \mu^+ \mu^-)$. An
important difference is however that the hard scale in $R$ is timelike
and requires one to stay out of the resonance mass region. In
contrast, $F_{\pi\gamma^*}$ involves spacelike virtualities and thus
offers the possibility to explore the quality of a leading twist
approximation down to moderate values of the hard scale. In that
respect it is similar to a number of sum rules in inclusive deep
inelastic scattering.  The fact that (\ref{qcd-pre}) should hold in a
wide region of $\omega$ raises hope for experimental feasibility of
this prediction.

One may ponder on whether at large $\qqb$ the form factor
$F_{\pi\gamma^*}$ has the potential for a determination of $\als$. Our
arguments in the appendix suggest that the suppression of $B_n$ by
$\omega^n$ holds to all orders in perturbation theory.  Higher orders
$\als$ coefficients in (\ref{qcd-pre}) for $\Phi_\pi=\Phi_{\rm AS}$
have been obtained in \cite{mue97} using the conformal operator
product expansion.  Given phenomenological or theoretical input on
$B_n$, the $\omega^n$ terms could at least be estimated, as could be
the size of power corrections.  Experimentally, the measurement of
$F_{\pi\gamma^*}$ should be quite clean. We will however see that
truly large $\qqb$ are not attainable at present facilities because of
rather small cross sections.

%%%%%%%%%%%%%%%%%%%%%%%%%%%%%%%%%%%%%%%%%%%%%%%%%%%%%%%%%%%%%%%%%%%%%%%
\section{The $\gamma^*$--$\,\eta$ and $\gamma^*$--$\,\eta'$ transition
form factors}
\label{sec:eta}
%%%%%%%%%%%%%%%%%%%%%%%%%%%%%%%%%%%%%%%%%%%%%%%%%%%%%%%%%%%%%%%%%%%%%%%

\begin{figure}
\parbox{\textwidth}{
\begin{center}
\psfig{file=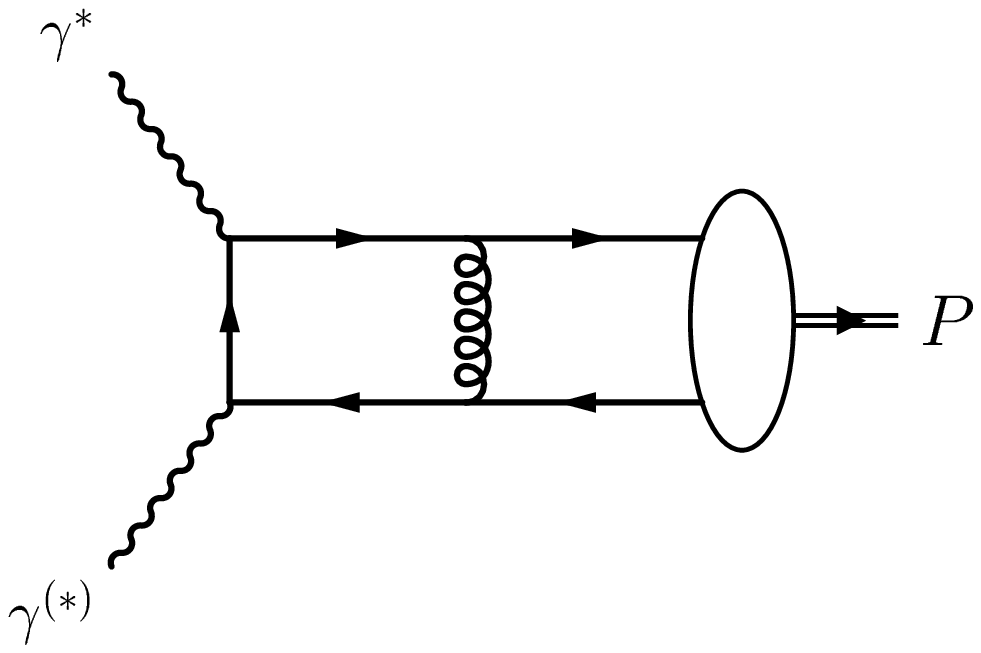,width=6.5cm} 
\hspace{4em}
\psfig{file=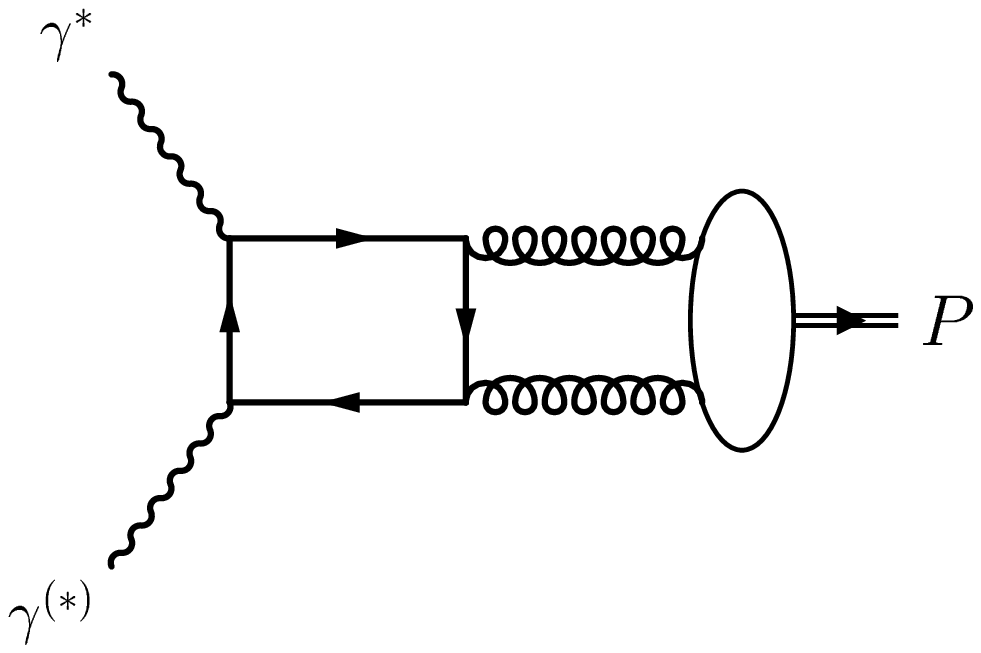,width=6.5cm}
\end{center}
}
\caption{Sample NLO Feynman graphs for the $\gamma^* \gamma^{(*)} \to
P$ transition.}
\label{fig:nlo}
\end{figure}

Let us now discuss the $\gamma^*$--$P$ transition form factors for $P
= \eta,\eta'$.  For the valence Fock states of the mesons we choose a
basis with the SU(3)$_{\rm F}$ singlet $|\qbq^{\,(1)}\rangle = |\ubu +
\dbd + \sbs\rangle/\sqrt{3}$ and octet $|\qbq^{\,(8)}\rangle = |\ubu +
\dbd - 2\sbs\rangle/\sqrt{6}$, and the two-gluon state
$|\glg\rangle$. This has the advantage that the corresponding \da{}s
$\Phi_P^{(1)}$ and $\Phi_P^{(g)}$ mix under evolution, but
$\Phi_P^{(8)}$ evolves independently. The solution of the LO evolution
equation for the octet \da{} is given by (\ref{evoleq}) with
Gegenbauer coefficients $B_{Pn}^{(8)}$, whereas for the quark singlet
and gluon one can write~\cite{ohr81}
\begin{eqnarray} 
\Phi_P^{(1)}(\xi,\mu_F)&=&\Phi_{\rm AS}(\xi) \left[1+ 
     \sum^\infty_{n=2,4,\ldots} B_{Pn}^{(1)} \,C_n^{3/2}(\xi)\right] ,
\nonumber\\
\Phi_P^{(g)}(\xi,\mu_F)&=&\frac{(1-\xi^2)^2}{16}
     \sum^\infty_{n=2,4,\ldots} B_{Pn}^{(g)} \,C_{n-1}^{5/2}(\xi) ,
\label{qda}
\end{eqnarray}
with
\begin{equation}
B_{Pn}^{(1)}(\mu_F) = B_{Pn}^{(+)}(\mu_F) + B_{Pn}^{(-)}(\mu_F) \,,
\hspace{2em}
B_{Pn}^{(g)}(\mu_F) = 
   a^{(+)}_{n \phantom{P}}\, B_{Pn}^{(+)}(\mu_F) +
   a^{(-)}_{n \phantom{P}}\, B_{Pn}^{(-)}(\mu_F) ,
\end{equation}
where the $B_{Pn}^{(\pm)}$ evolve as in (\ref{evolve}) with positive
anomalous dimensions $\gamma_n^{(\pm)}$. We remark in passing that
conflicting results on $\gamma_n^{(\pm)}$ and $a_n^{(\pm)}$ are found
in the literature~\cite{ohr81}.  Notice that the two-gluon \da{}
vanishes in the asymptotic limit $\ln\mu_F \to\infty$. It contributes
to $F_{P\gamma^*}$ only to order $\als$ through the box graph shown in
Fig.\ \ref{fig:nlo}. The corresponding hard scattering amplitude can
be obtained by crossing from the NLO corrections to the Compton
amplitude $\gamma^*p\to \gamma^*p$, which can be found in
\cite{man98}.

In full analogy to the case of the $\pi$ we can write the
$\gamma^*$--$P$ transition form factor as a superposition of
Gegenbauer coefficients with $\omega$ dependent weights. In the limit
$\omega\to 1$ one probes essentially the sums $\sum_n B_{Pn}^{(8)}$
and $\sum_n B_{Pn}^{(1)}$ of coefficients appearing in the quark
\da{}s $\Phi^{(8)}_P$ and $\Phi^{(1)}_P$.  The real-photon limit has
been analysed in \cite{jak96,cao96,fel97}.  Using the measurements of
CLEO \cite{cleo98} and L3 \cite{acc98}, it was found in
Ref.~\cite{fel97} that within experimental accuracy the data on
$F_{\eta\gamma}$ and $F_{\eta'\gamma}$ is compatible with the
asymptotic forms of the quark distribution amplitudes and
correspondingly vanishing gluon ones.

When $\omega$ moves away from 1 the form factors become increasingly
less sensitive to the higher order coefficients.  As in the case of
quarks, we find that the Gegenbauer coefficient $B^{(g)}_{Pn}$ or the
gluon \da{} first appears at order $\omega^n$, despite the fact that
$\Phi_P^{(g)}$ is expanded upon Gegenbauer polynomials $C_{n-1}^{5/2}$
instead of $C_{n \phantom{1}}^{3/2}$. In analogy to~(\ref{qcd-pre}) we
then obtain the prediction
\begin{equation}
F_{P\gamma^*}(\qb,\omega) = 
   \frac{\sqrt{2}\: f_P^{{\rm eff}}}{3\: \qqb}
   \, \bigg[ 1 - \frac{\als}{\pi} \bigg ] 
+ {\cal O}(\omega^2 ,\als^2) \,.
\label{qcd-pre-eta}
\end{equation}
The effective (process-dependent) decay constants are
\begin{eqnarray}
f_P^{{\rm eff}} = 
  \frac{f_P^{(8)} + 2\sqrt{2} f_P^{(1)}}{\sqrt{3}} \,,
\end{eqnarray}
where 
\begin{equation}
       \langle 0 \mid J^{(i)}_{5\, \mu} \mid P(p) \rangle = i
                      f_P^{(i)}\, p_\mu ,   \qquad (i=1,8)
\label{dec-def}
\end{equation}
are matrix elements of the SU(3)$_{\rm F}$ singlet or octet axial
vector currents.

The decomposition of the mesons states $|\eta\rangle$ and
$|\eta'\rangle$ on $|\qbq^{\,(1)}\rangle$, $|\qbq^{\,(8)}\rangle$,
$|\glg\rangle$ and higher Fock states is completely general. It does
not refer to $\eta -\eta'$ mixing, which is known to occur
empirically, and which relates the respective Fock state coefficients
for the two mesons. In \cite{fel98,fel99} a quark-flavor mixing scheme
has been proposed, which successfully describes many physical
processes. In this scheme the physical mesons are obtained from two
basis states, $|\eta_q\rangle$ and $|\eta_s\rangle$ by a unitary
transformation with a mixing angle $\varphi$.  The quark valence Fock
states of $|\eta_q\rangle$ and $|\eta_s\rangle$ respectively are
$|\ubu + \dbd\rangle/\sqrt{2}$ and $|\sbs\rangle$.    
This mixing scheme can be justified to the extent that the \da{}s are
close to the asymptotic form; evolution does then practically not
spoil this scheme.  The decay constants (\ref{dec-def}) are given here
by the mixing angle and the two basic decay constants, $f_q$ and
$f_s$. For the effective decay constants above one then finds
\cite{fel98}
\begin{eqnarray}
    f_{\eta\phantom{'}}^{{\rm eff}} &=& \frac{5 f_q \cos{\varphi}
                    -\sqrt{2} f_s \sin{\varphi}}{3} \,, 
\nonumber\\ 
    f_{\eta'}^{{\rm eff}} &=& \frac{5 f_q \sin{\varphi}
                    +\sqrt{2} f_s \cos{\varphi}}{3} \,.
\end{eqnarray}
The phenomenological values of the mixing parameters derived in
\cite{fel98} numerically give $f_{\eta}^{{\rm eff}}=0.98 f_\pi$ and
$f_{\eta'}^{{\rm eff}}= 1.62 f_\pi$. Similar results are obtained with
the large $N_c$ values of the $f_P^{(i)}$~\cite{kai98}. At small
$\omega$ and large enough $\qqb$ the ratio of the $\gamma^*$--$\,\eta$
and $\gamma^*$--$\,\eta'$ form factors constitutes an accurate measure
of the effective decay constants. This can be used for a severe test
of the $\eta - \eta'$ mixing scheme. As in the pion case, substantial
deviations from the small-$\omega$ predictions (\ref{qcd-pre-eta})
only occur if there are unexpectedly large power corrections or
extremely large Gegenbauer coefficients.

One may extend our analysis to the gluonic transitions $g^* g^* \to
\eta, \eta'$ in the same fashion as we discussed the electromagnetic
ones.  The case of the $ \eta'$ has recently been investigated by Ali
and Parkhomenko \cite{ali2000}, and we can follow their analysis
closely.  For $\omega \to 1$ the form factors not only depend on the
Gegenbauer coefficients $B^{(1,8)}_{P,n}$ but are also sensitive to
variations of the $B^{(g)}_{Pn}$, since in contrast to the
electromagnetic case both quark and gluon \da{}s now contribute to LO
in $\als$.  On the other hand, we can still apply our arguments for
the suppression of the Gegenbauer coefficients (and hence of the full
contribution from $\Phi^{(g)}_P$) in the limit $\omega \to 0$. We find
for the transition form factors
\begin{equation}
F_{P g^*}(\qb,\omega) = - \frac{4\pi\als}{3\, \qqb}\, C_{P}  
   + {\cal O}(\omega^2,\als^2)
\end{equation}
with effective decay constants
\begin{eqnarray}
C_\eta & =& \sqrt{2} f_q \cos {\varphi} - f_s \sin{\varphi} 
   \approx 0.32 f_\pi\,, 
\nonumber\\  
C_{\eta'}&=&\sqrt{2}\, f_q \sin{\varphi}+f_s \cos{\varphi} 
   \approx 1.99 f_\pi\,. 
\end{eqnarray}
The ratio of the two form factors at small $\omega$ is given by
$C_\eta/C_{\eta'}=-\tan{\theta_1}$ which, if measurable, would give
access to the badly constrained $\eta-\eta'$ mixing parameter
$\theta_1$ \cite{fel99}. This mixing angle determines the relative
decay strength of the $\eta$ and $\eta'$ through a weak SU(3)$_{\rm
F}$ singlet current. Present estimates of the angle $\theta_1$ range
from $-2^\circ$ to $-10^\circ$ \cite{fel99}.

%%%%%%%%%%%%%%%%%%%%%%%%%%%%%%%%%%%%%%%%%%%%%%%%%%%%%%%%%%%%%%%%%%%%%%
\section{Cross section estimates}
\label{sec:cross}
%%%%%%%%%%%%%%%%%%%%%%%%%%%%%%%%%%%%%%%%%%%%%%%%%%%%%%%%%%%%%%%%%%%%%%

In order to assess the possibilities of the running experiments BaBar,
Belle, and CLEO to investigate $F_{\pi\gamma^*}$ we will now estimate
cross sections and see how they are affected by acceptance cuts. We
first remark that at large $\qqb$ the cross section for $e^+ e^- \to
e^+ e^- \, \pi^0$ depends on the two photon virtualities roughly like
\begin{equation}
\frac{d\sigma}{dQ^2\, dQ'^2} \sim 
      \frac{1}{Q^2\, Q'^2\, (Q^2+Q'^2)^2} ,
\label{X-section}
\end{equation}
where the factors $Q^2$ and $Q'^2$ are due to the photon flux, and
$(Q^2+Q'^2)^2$ comes from the behavior of $F_{\pi\gamma^*}$, which as
we have seen behaves like $1/(Q^2+Q'^2)$ with only a mild dependence
on $\omega$.

For acceptance cuts we will consider two scenarios. One is that both
scattered leptons and the $\pi^0$ are seen in the detector. In
kinematics where this is not possible because either $Q^2$ or $Q'^2$
is too small to ensure a sufficiently large lepton scattering angle,
one may envisage the detection only of one lepton and the $\pi^0$. If
experimental resolution permits, the four-momentum of the undetected
lepton can then be reconstructed using four-momentum conservation.

We define polar angles with respect to the direction of the
\emph{positron} beam, and take the convention that $Q^2$ corresponds
to the photon radiated from the $e^+$. As typical cuts we impose
for tagged leptons a minimum energy of 1~GeV and transverse momentum
of 100~MeV, and a minimum transverse momentum of 200~MeV for the
pion. We further demand that a tagged $e^+$ has a polar angle
$\alpha_{e^+} > \alpha_{\mathrm{min}}$ in the detector. Likewise, we
require $\alpha_{e^-} < \alpha_{\mathrm{max}}$ for a tagged $e^-$, and
$\vartheta_{\mathrm{min}} < \vartheta_\pi < \vartheta_{\mathrm{max}}$
for the polar angle $\vartheta_\pi$ of the pion. For the minimal and
maximal angles we take the values given in Table~\ref{tab:cuts}. To
estimate $F_{\gamma^*\pi}$ we use the leading-twist NLO expression
(\ref{fpgvirtual}), taking for simplicity the asymptotic form
$\Phi_{\rm AS}$ of the pion \da.

\begin{table}
\begin{center}
\begin{tabular}{lcccccc}  \hline\hline
 & $E_{e^+}$ & $E_{e^-}$ & 
   $\alpha_{\mathrm{min}}$    & $\pi-\alpha_{\mathrm{max}}$ & 
   $\vartheta_{\mathrm{min}}$ & $\pi-\vartheta_{\mathrm{max}}$ \\
\hline 
BaBar & 3.1 & 9.0 &  478 & 284  & 667 & 275 \\
Belle & 3.5 & 8.0 &  154 & 112  & 524 & 294 \\
CLEO  & 5.3 & 5.3 &  227 & 227  & 314 & 314 \\
\hline\hline
\end{tabular}
\end{center}
\caption{\label{tab:cuts}Minimal and maximal values of polar angles in
the laboratory frame imposed by our cuts, as explained in the text. We
also give the beam energies for the different experiments.  Angles
refer to the positron beam axis and are given in mrad, energies are
given in GeV.}
\end{table}

\begin{figure}
\begin{center}
\leavevmode
\epsfxsize=0.49\textwidth  \epsffile{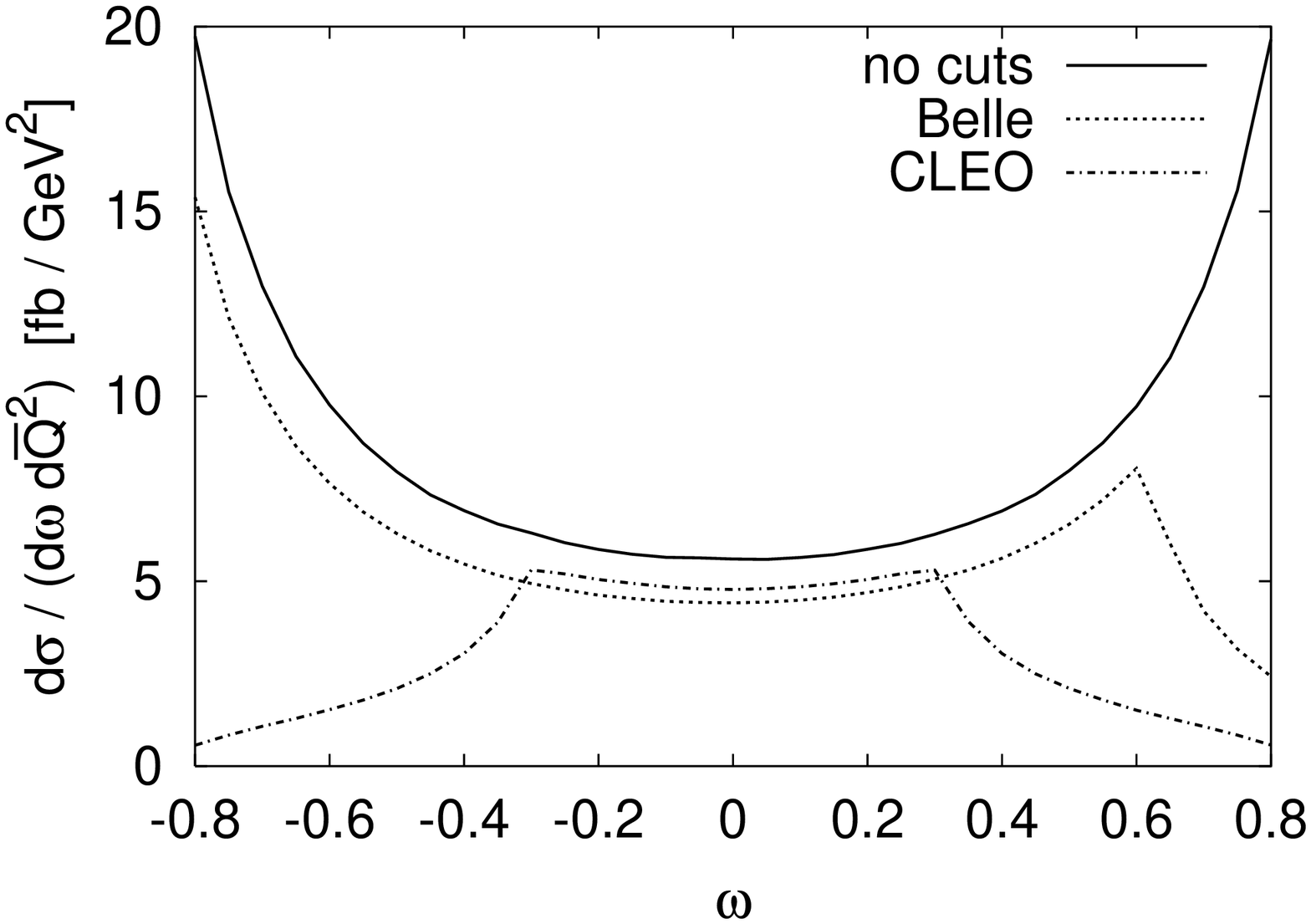}  
\hfill
\epsfxsize=0.49\textwidth  \epsffile{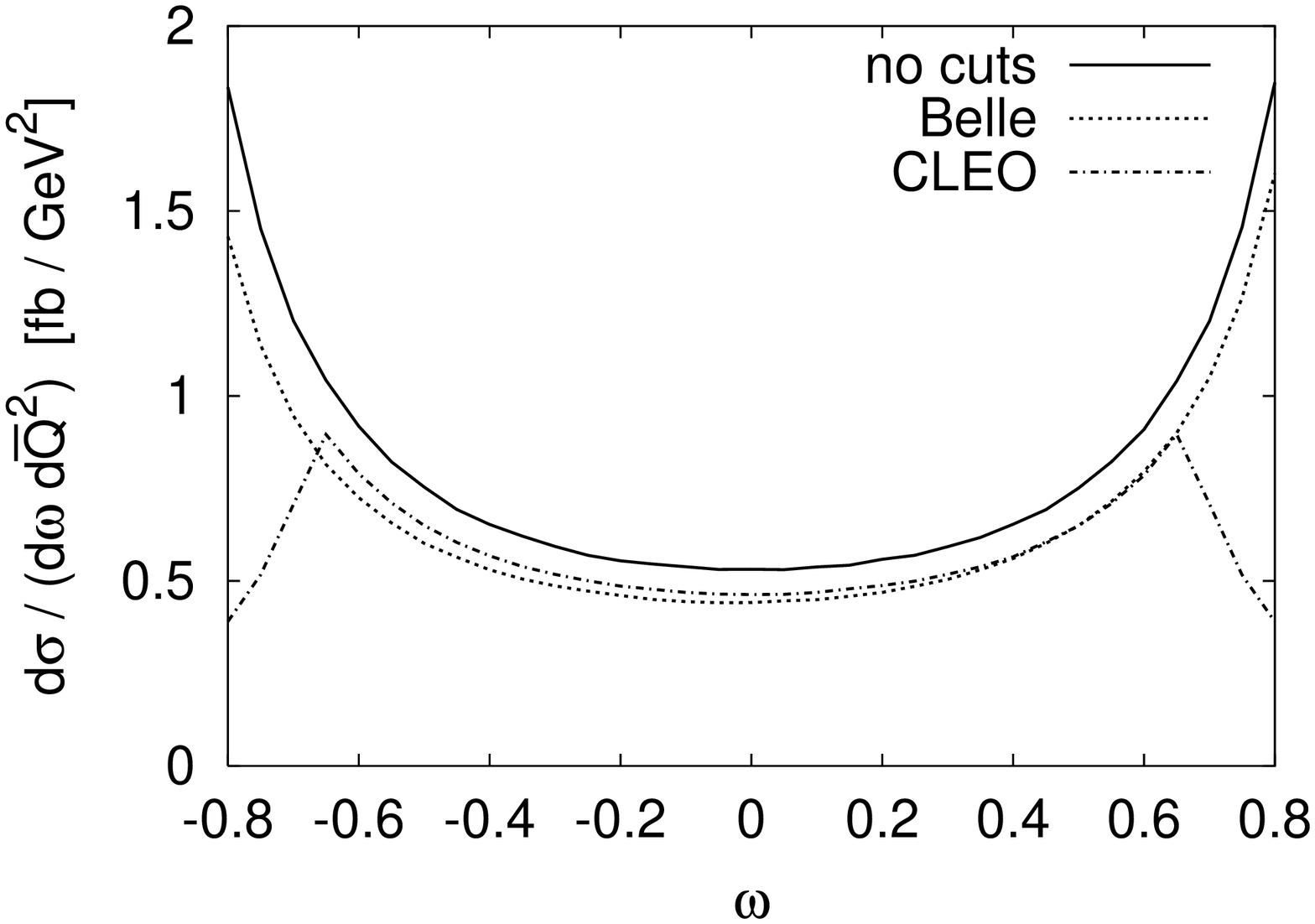}
%\end{center}
\caption[]{\label{fig:cent} The differential cross section for
$e^+e^-\to e^+e^-\pi$ as a function of $\omega$ for $\qqb = 2\gev^2$
(left) and $\qqb = 4\gev^2$ (right). Notice that the scaling behavior
(\protect\ref{X-section}) translates into $d\sigma /(d\omega\,
d\qb{}^{\,2}) \sim (1-\omega^2)^{-1} \,\qb{}^{-6}$. }
%\end{figure}
%\begin{figure}[h]
%\begin{center}
\vspace{1cm}
\leavevmode
\epsfxsize=0.49\textwidth  \epsffile{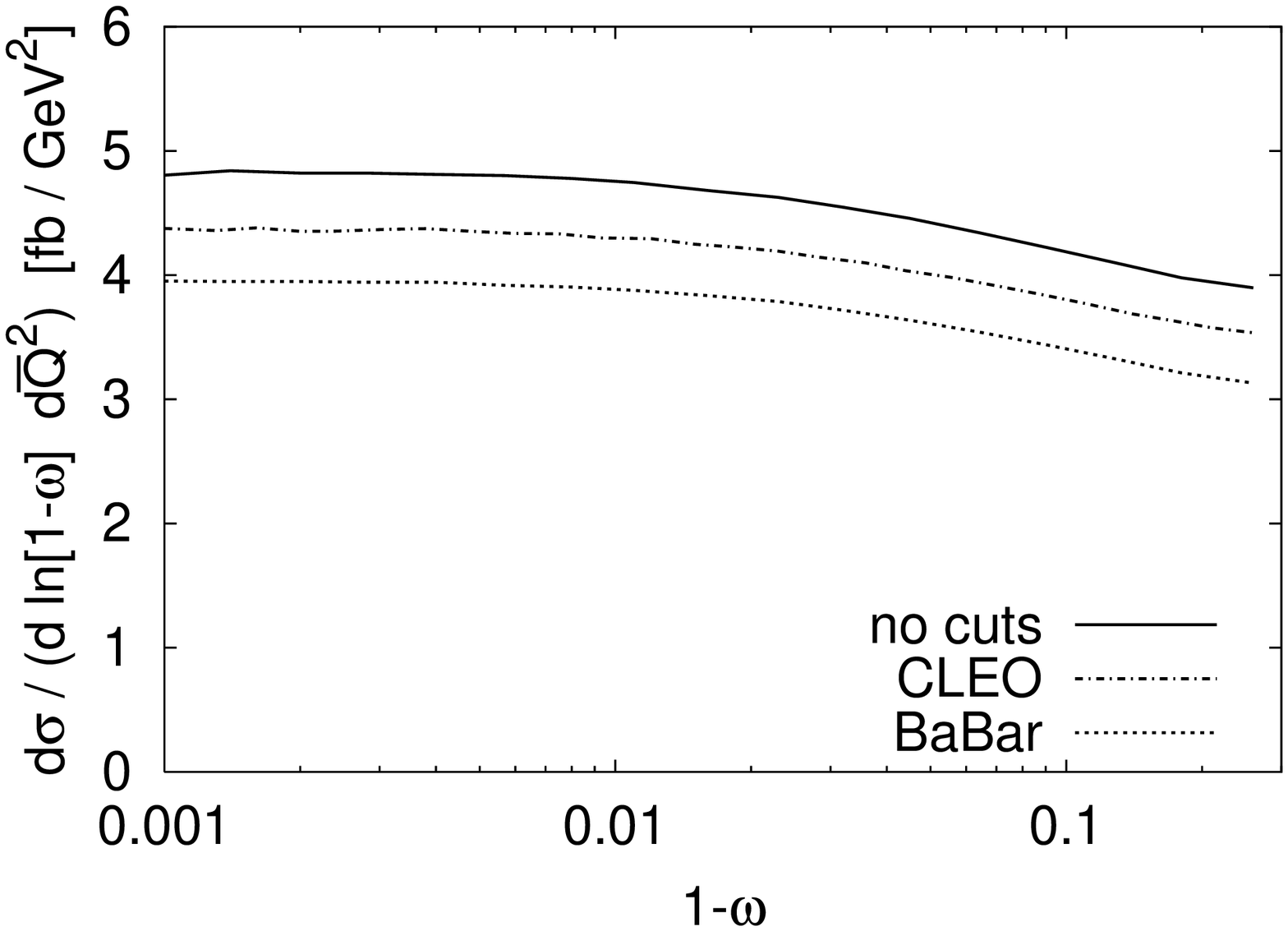}
\hfill
\epsfxsize=0.49\textwidth  \epsffile{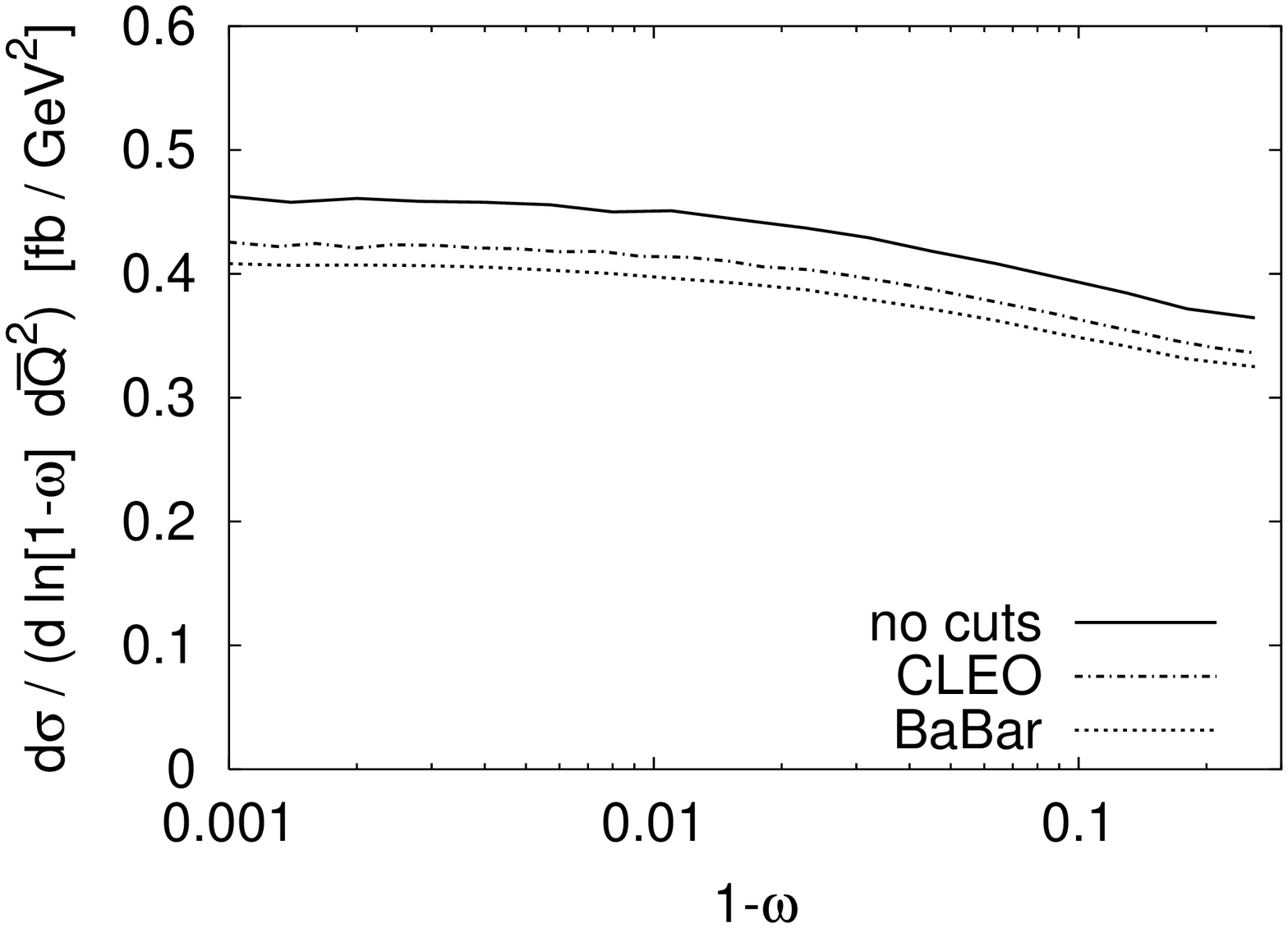}
\end{center}
\caption[]{The differential cross section near $\omega=1$ for
$\qqb=2\gev^2$ (left) and $\qqb = 4\gev^2$ (right). Corresponding
curves with cuts for Belle are between those for BaBar and for
CLEO. In this region, the behavior (\protect\ref{X-section}) becomes
$d\sigma /(d\log[1-\omega]\, d\qb{}^2) \sim \qb{}^{-6}$.}
\label{fig:rim}
\end{figure}

It turns out that the cuts with the most serious impact are the
angular restrictions on the scattered leptons, while the precise
values of the other cuts have only a mild influence. We show in
Fig.~\ref{fig:cent} the differential $e^+e^-$ cross section in the
region $Q^2\sim Q'^2$. For Belle and CLEO we have imposed detection
cuts for both scattered leptons. One clearly sees how for larger
$|\omega|$, where at fixed $\qqb$ one of the photon virtualities
becomes small, our cuts do have visible effects. Imposing the same
cuts for BaBar leaves essentially no cross section in the kinematical
region we are considering. This is due to the limited forward and
backward coverage of the BaBar detector: for photon virtualities large
enough to bring the scattered lepton within detector acceptance, the
cross section is already minute because of its strong decrease
(\ref{X-section}) with $Q^2$ and $Q'^2$. The same holds if we allow
the $e^+$ to be undetected. Only in the case where the $e^-$ is
untagged do we obtain a signal, not shown in
Fig.~\ref{fig:cent}.

In Fig.~\ref{fig:rim} we show the differential cross section in the
region of $\omega$ close to 1, requiring only the scattered $e^+$ and
the $\pi^0$ to be observed. If instead we require the $e^-$ and
$\pi^0$ to be detected in the region of $\omega$ close to $-1$, we
obtain somewhat smaller cross sections in the case of Belle. For BaBar
we have little change at $\qqb=4 \gev^2$ but almost no signal left at
$\qqb=2 \gev^2$. For CLEO with its symmetric geometry, there is of
course no difference between the two cases.

We have also investigated the production of an $\eta$ or $\eta'$,
imposing the same cuts as described for the pion case. The
corresponding cross sections scale approximately like the squared
transition form factors $F_{P\gamma^*}$, i.e., the mass differences
between the $\pi$, $\eta$, and $\eta'$ have only little effect on
kinematics and phase space in the region we are investigating.

Concerning the values of cross section estimated here, we recall the
benchmark luminosity of 30~fb$^{-1}$ per year of the $B$
factories. With the numbers in Fig.~\ref{fig:cent} and \ref{fig:rim}
we conclude that for $\qqb$ around 2~GeV$^2$ studies should be
possible, both when $\omega$ is around zero and when $|\omega| \approx
1$. As $\qqb$ goes up to 4~GeV$^2$ and more, event statistics will
become increasingly problematic, so that, unfortunately, we do not
expect precision measurements to feasible in that region with the
current experimental setups. Such studies would greatly benefit from
high-luminosity upgrades of the $B$-factories.

We remark that our rate estimates here are restricted to $|\omega| \le
1 - 10^{-3}$ and thus do not include the real-photon limit. There the
cross section will be much higher and, as emphasized at the end of
Sect.~\ref{sec:real}, measurements with good statistics should be
possible at higher values of $Q^2$.

%%%%%%%%%%%%%%%%%%%%%%%%%%%%%%%%%%%%%%%%%%%%%%%%%%%%%%%%%%%%%%%%%%%%%%
\section{Summary}
\label{sec:sum}
%%%%%%%%%%%%%%%%%%%%%%%%%%%%%%%%%%%%%%%%%%%%%%%%%%%%%%%%%%%%%%%%%%%%%%

We have investigated the $\gamma^*$--$\,\pi$ transition form factor to
leading-twist accuracy including $\als$ corrections. The chief purpose
of our analysis is to assess what can be learned about the pion \da{}
from experimental data on $F_{\pi\gamma^*}$.

Our main idea is to use the expansion of $\Phi_\pi$ on Gegenbauer
polynomials, and to write $F_{\pi\gamma^*}$ as a double series in the
Gegenbauer coefficients $B_n$ and in powers of the variable $\omega$,
which describes the difference of the two photon virtualities. We find
that, contrary to what one may expect, it is very difficult to obtain
information on the $B_n$ beyond what can be inferred from the case
where one of the photons is quasi-real, which essentially constrains
the sum of Gegenbauer coefficients. Only for $\omega$ values close to
but less than 1 can one get more information. Effects on
$F_{\pi\gamma^*}$ of the order of 10\% to 15\% can be obtained using
coefficients whose magnitude is not implausibly large compared to
theory estimates. Data in that range of $\omega$ can, for instance,
allow a check whether the small value of the sum $\sum_n B_n$,
extracted from the CLEO real-photon data, results from cancellations
of rather large individual terms or from the smallness of the $B_n$
themselves.  This type of information would be a valuable input into
other phenomenological studies of $\Phi_\pi$.

One of our main findings is that a Gegenbauer coefficient $B_n$ in
$\Phi_\pi$ contributes to $F_{\pi\gamma^*}$ with a weight proportional
to $\omega^{n}$.  For a large range of $\omega$, in fact for $\omega
\lsim 0.6$, we find that the form factor is independent of the
Gegenbauer coefficients to a high degree of accuracy.  Although this
is bad news for a determination of $\Phi_\pi$, it entails a parameter
free QCD prediction of the $\gamma^* \gamma^* \to\pi$ form factor. Any
clear deviation from this result observed in experiment would be an
unambiguous signal for power corrections, provided that the lowest
Gegenbauer coefficients $B_n$ are not extremely large, a scenario
which could be ruled out using the region $\omega \simeq 1$.  In a
wide region of $\omega$ around zero, data on $F_{\pi\gamma^*}$ would
thus permit one to test the quality of leading twist QCD in the
spacelike region, in a similar fashion as sum rules in deep inelastic
scattering.

Analogous results hold for the $\gamma^* \gamma^* \to\eta$ and
$\gamma^* \gamma^* \to\eta'$ form factors. Note that here the gluon
\da{}s contribute indirectly through mixing in the evolution and
directly to order $\als$. Their contribution is further suppressed for
small $\omega$. Data on $F_{\eta\gamma^*}$ and $F_{\eta'\gamma^*}$ in
that region could rather cleanly determine of a linear combination of
the flavor singlet and octet decay constants of these mesons, and thus
give valuable information on $\eta$--$\eta'$ mixing.

Cross section estimates of the process $e^+ e^- \to e^+ e^- \pi^0$ at
the running experiments BaBar, Belle, and CLEO indicate that it should
be possible, although challenging, to measure the transition form
factors for virtual photons up to about $\qqb \lsim 4
\gev^2$. Limiting factors for these measurements are luminosity and
the acceptance for lepton detection in the forward and backward
regions of the detector. Measurements with better statistics and at
higher $\qqb$ might be feasible at high-luminosity upgrades of the
present $B$-factories.

Concerning the real-photon limit, we argue that although the present
data on $F_{P\gamma}$ favor a small value for the sum $\sum_n B_n$,
more precise statements can only be made at the price of theory
assumptions on the nature and size of power corrections to the
leading-twist result. High statistics measurements at large
virtualities $Q^2$ would greatly alleviate this problem and should be
feasible at BaBar, Belle, and CLEO. They should be worthwhile, since
the pseudoscalar meson \da{}s are fundamental quantities describing
meson structure and providing benchmark tests for nonperturbative
methods in QCD. They are also an input required for the calculation of
several phenomenologically important processes in hard-scattering
approaches. An example are exclusive nonleptonic $B$ meson decays into
pseudoscalars, where a good understanding of the strong interaction
dynamics would enhance the prospects of extracting information on
$C\!P$ violation.

%%%%%%%%%%%%%%%%%%%%%%%%%%%%%%%%%%%%%%%%%%%%%%%%%%%%%%%%%%%%%%%%%%%%%%
\section*{Acknowledgments} 
%%%%%%%%%%%%%%%%%%%%%%%%%%%%%%%%%%%%%%%%%%%%%%%%%%%%%%%%%%%%%%%%%%%%%%

We wish to acknowledge discussion with A. Ali, J.~C.~Collins,
T. Feldmann, A. Grozin, R. Jakob, A. Khodjamirian, H. Koch,
D. M\"uller and V. Savinov. C.~V. thanks the Deutsche
Forschungsgemeinschaft for support.

%%%%%%%%%%%%%%%%%%%%%%%%%%%%%%%%%%%%%%%%%%%%%%%%%%%%%%%%%%%%%%%%%%%%%%
\begin{appendix}
\section*{Appendix}
%%%%%%%%%%%%%%%%%%%%%%%%%%%%%%%%%%%%%%%%%%%%%%%%%%%%%%%%%%%%%%%%%%%%%%

The NLO hard scattering kernel in (\ref{fpgvirtual}), evaluated in the
$\overline{\rm MS}$ scheme, reads
\begin{eqnarray}
\lefteqn{ {\cal K} = \frac{1}{6}\, \Big[ \,
  (1+\xi\omega) \ln(1-\xi\omega) + 4 (1-\omega) \ln(1-\omega) }
\nonumber \\
&& \hspace{0.8em} {}
   + (1+\xi\omega) \ln^2(1-\xi\omega) 
   - (1-\omega) \ln^2(1-\omega) - 9 (1+\xi\omega)  \,\Big]
\phantom{\frac{1}{2}}
\nonumber \\
&& {}+ \frac{1}{6}\, \ln\frac{\qqb}{\mu_F^2} \Big[ \,
    2 (1+\xi\omega) \ln(1-\xi\omega) - 2 (1-\omega) \ln(1-\omega)
  + 3 (1+\xi\omega) \Big]
\nonumber \\
&& {}+ \frac{1}{6\,\omega^2(1-\xi^2)}  \Big[ \,
    2 (1+\xi\omega) (1+\xi\omega-2\,\omega^2) \ln(1-\xi\omega)
  - 2 (1+\omega) (1+\omega-2\,\omega^2) \ln(1-\omega)  
\nonumber \\
&& \hspace{5.9em} {}
  - (1+\xi\omega) (1-\omega^2) \ln^2(1-\xi\omega)
  + (1+\omega) (1-\omega^2) \ln^2(1-\omega) \,\Big]  
\phantom{\frac{1}{2}}
\nonumber \\
&& {}- \frac{1-\omega^2}{3\,\omega^2(1-\xi^2)}\, 
       \ln\frac{\qqb}{\mu_F^2} \Big[ \,
    (1+\xi\omega) \ln(1-\xi\omega) - (1+\omega) \ln(1-\omega)  \,\Big]
\nonumber \\
&& {}+ \{ \omega \to -\omega \}  \,.
\phantom{\frac{1}{2}} 
\label{nlo-kernel}
\end{eqnarray}
We will now show that the relevant convolution in the NLO part of
$F_{\pi\gamma^*}$ can be written as
\begin{eqnarray}
\lefteqn{ \int_{-1}^{1} d\xi\, \frac{1-\xi^2}{1-\xi^2\omega^2} 
 \, {\cal K}(\omega,\xi)
 \left[1+ \sum^\infty_{n=2,4,\ldots} B_n \,C_n^{3/2}(\xi)\right] }
\nonumber \\
&=& \sum^\infty_{m=0,2,\ldots}  \omega^m\, 
                \int_{-1}^{1} d\xi\, (1-\xi^2)\, p_m (\xi)
+ \sum^\infty_{\scriptstyle n=2,4,\ldots 
           \atop \scriptstyle m=n,n+2,\ldots}  \omega^m B_n \,
    \int_{-1}^{1} d\xi\, (1-\xi^2)\, p_m(\xi)\, 
    C_n^{3/2}(\xi) \,,
\hspace{1em}
 \label{omega-limit-nlo}
\end{eqnarray}
where the $p_m(\xi)$ are polynomials in $\xi$ of order $m$.  Due to
the orthogonality of the $C_n^{3/2}(\xi)$ the sum involving the $B_n$
is here again restricted to $m\ge n$. To show (\ref{omega-limit-nlo})
it is enough to establish that ${\cal K}(\omega,\xi)$ can be expanded
in a double Taylor series
\begin{equation}
\sum_{m=0}^\infty \sum_{l=0}^m d_{m-l,l}\, \omega^{2m} \xi^{2l} =
\sum_{k,l=0}^\infty d_{k,l}\, \omega^{2k} (\omega\xi)^{2l} \,,
\end{equation}
i.e., that ${\cal K}$ is analytic in the two variables $\omega^2$ and
$\omega^2\xi^2$ at $\omega^2=\omega^2\xi^2=0$. One readily sees that
${\cal K}$ is even in $\omega$ and $\xi$, which is a consequence of
Bose symmetry and charge conjugation invariance. Further inspection of
(\ref{nlo-kernel}) shows that we can write
\begin{equation}
{\cal K} = f(\omega^2, \omega^2\xi^2)
   + \frac{g(\omega^2, \omega^2\xi^2)}{\omega^2 - \omega^2\xi^2}
\label{structure}
\end{equation}
where $f$ and $g$ are analytic in their variables around
$\omega^2=\omega^2\xi^2=0$. Hence $g$ is also analytic in $\omega^2$
and $\omega^2 - \omega^2\xi^2$. Finally one can see from
(\ref{nlo-kernel}) that $g$ is zero for $\xi=1$. The apparent pole at
$\omega^2 - \omega^2\xi^2=0$ thus cancels in (\ref{structure}), and
$g /(\omega^2 - \omega^2\xi^2)$ also has the required analyticity
properties.

It is not surprising that ${\cal K}$ is analytic in the pair of
variables $\omega$ and $\omega\xi$. Apart from the appropriate
subtraction of collinear singularities, the NLO hard scattering kernel
is the amplitude for the partonic subprocess $\gamma^*(q) +
\gamma^*(q')\to q(k) + \bar{q}(k')$ in the collinear limit, i.e., at
$(k+k')^2=0$. {}From the relations $q'^2 = - \qqb (1-\omega)$ and
$(q'-k')^2 = - \qqb (1 - \omega\xi)$ we see that for $\qqb > 0$
analyticity in $\omega$ and $\omega\xi$ around $\omega=\omega\xi=0$ is
equivalent to analyticity in the spacelike invariants $q'^2$ and
$(q'-k')^2$. This suggests that our result (\ref{omega-limit}) will
generalize to higher orders in $\als$, provided appropriate
analyticity properties of the collinear subtraction terms in the hard
scattering kernel.

\end{appendix}

%%%%%%%%%%%%%%%%%%%%%%%%%%%%%%%%%%%%%%%%%%%%%%%%%%%%%%%%%%%%%%%%%%%%%%

\end{document}